\documentclass[12pt]{article}
\usepackage[margin=1in]{geometry}
\usepackage[utf8]{inputenc}
\usepackage{amsmath,amssymb,bm,url,mathrsfs}
\usepackage{algorithmic}
\usepackage[ruled,linesnumbered]{algorithm2e}
\usepackage{amsthm}
\usepackage{hyperref}
\usepackage{makecell}
\usepackage{tablefootnote}
\usepackage{comment}
\usepackage{soul}
\usepackage{graphics,graphicx}
\usepackage{subcaption,caption,cleveref}
\hypersetup{
    colorlinks=true,
    linkcolor=blue,
    citecolor=blue,
    filecolor=magenta,      
    urlcolor=cyan,
}
\usepackage{booktabs}
\usepackage{multirow}
\usepackage{multicol}
\usepackage{natbib}
\usepackage{xcolor}
\usepackage{dsfont}
\usepackage{bbm}
\usepackage{enumitem}
\usepackage{blkarray}
\usepackage{amsfonts}
\usepackage{accents}

\def\frakE{{\mathfrak E}}

\def\frakG{{\mathfrak G}}

\def\frakS{{\mathfrak S}}

\def\calE{{\mathcal E}}

\def\calP{{\mathcal P}}

\def\calS{{\mathcal S}}

\def\calV{{\mathcal V}}

\def\EE{{\mathbb E}}

\def\II{{\mathbb I}}

\def\PP{{\mathbb P}}

\def\RR{{\mathbb R}}

\def\a{{\boldsymbol a}}
\def\b{{\boldsymbol b}}

\def\e{{\boldsymbol e}}

\def\p{{\boldsymbol p}}

\def\s{{\boldsymbol s}}

\def\u{{\boldsymbol u}}
\def\v{{\boldsymbol v}}

\DeclareMathOperator*{\argmax}{arg\,max}
\DeclareMathOperator*{\argmin}{arg\,min}

\def\bs{\boldsymbol{s}}

\def\calE{{\cal  E}}

\def\calP{{\cal  P}}

\def\calS{{\cal  S}}

\def\calV{{\cal  V}}

\def\bzero{\bfm 0}

\newcommand{\bfm}[1]{\ensuremath{\mathbf{#1}}}

   \def\bA{\bfm A}  
   \def\bB{\bfm B}  
\def\bc{\bfm c}     
     
   \def\bE{\bfm E}  \def\EE{\mathbb{E}}
    
   \def\bG{\bfm G}  
     
     \def\II{\mathbb{I}}

     \def\PP{\mathbb{P}}
     
   \def\bR{\bfm R}  \def\RR{\mathbb{R}}
\def\bs{\bfm s}     
     
   \def\bU{\bfm U}  
   \def\bV{\bfm V}

   \def\bZ{\bfm Z}  
                   
  \def\bTheta{\bfm \Theta}

\def\bSigma{\bfm \Sigma}

\allowdisplaybreaks

\def\hat{\widehat}
\def\wt{\widetilde}

\newcommand{\mb}{\mathbf }

\newcommand\fro[1]{\left\| #1 \right\|_{\rm F}}
\newcommand\op[1]{\left\|#1\right\|}

\newcommand\brac[1]{\left(#1\right)}
\newcommand\bbrac[1]{\Big(#1\Big)}
\newcommand\ebrac[1]{\left\{#1\right\}}
\newcommand\sqbrac[1]{\left[#1\right]}
\newcommand\ab[1]{\left|#1\right|}

\newtheorem{theorem}{Theorem}

\newtheorem{proposition}{Proposition}
\newtheorem{remark}{Remark}

\newtheorem{assumption}{Assumption}
\newtheorem{lemma}{Lemma}
\newtheorem{corollary}{Corollary}

\theoremstyle{remark}

\newcommand\munderbar[1]{%
  \underaccent{\bar}{#1}}

\usepackage{xr}
\def\spacingset#1{\renewcommand{\baselinestretch}%
{#1}\small\normalsize}
\spacingset{1}
\begin{document}

\title{Spectral Clustering with Likelihood Refinement  for High-dimensional Latent Class Recovery}

\author{Zhongyuan Lyu and Yuqi Gu\footnote{Correspondence to: Yuqi Gu, Department of Statistics, Columbia University. Email: \texttt{yuqi.gu@columbia.edu}. Address: Room 928, 1255 Amsterdam Avenue, New York, NY 10025.}}

\date{Columbia University}

\maketitle

\begin{abstract}
Latent class models are widely used for identifying unobserved subgroups from multivariate categorical data in social sciences, with binary data as a particularly popular example. However, accurately recovering individual latent class memberships remains challenging, especially when handling high-dimensional datasets with many items. This work proposes a novel two-stage algorithm for latent class models suited for high-dimensional binary responses. Our method first initializes latent class assignments by an easy-to-implement spectral clustering algorithm, and then refines these assignments with a one-step likelihood-based update. This approach combines the computational efficiency of spectral clustering with the improved statistical accuracy of likelihood-based estimation. We establish theoretical guarantees showing that this method is minimax-optimal for latent class recovery in the statistical decision theory sense. The method also leads to exact clustering of subjects with high probability under mild conditions. As a byproduct, we propose a computationally efficient consistent estimator for the number of latent classes. Extensive experiments on both simulated data and real data validate our theoretical results and demonstrate our method's superior performance over alternative methods.
\end{abstract}

\noindent\textbf{Keywords}: 
High-dimensional data;
Joint maximum likelihood estimation;
Latent class model; 
Spectral clustering.

\spacingset{1.6}
\section{Introduction}\label{sec:intro}
Latent class models \citep[LCMs,][]{lazarsfeld1968latent, goodman1974exploratory} are popular tools for uncovering unobserved subgroups  from multivariate categorical responses, with broad applications in social and behavioral sciences \citep{korpershoek2015differences, berzofsky2014local, wang2011latent, zeng2023tensor}. 
At the core of an LCM, we aim to identify $K$ latent classes of $N$ individuals based on their categorical responses to $J$ items. In this paper, we focus on the widely used LCMs with binary responses, which are prevailing in  educational assessments (correct/wrong responses to test questions) and social
science surveys (yes/no responses to survey items).
Accurately and efficiently recovering latent classes in large $N$ and large $J$ settings remains a significant statistical and computational challenge.

Traditional approaches to LCM are typically based on maximum likelihood estimation. In the classical random-effect formulation, the latent class labels of subjects are treated as random variables and the model is estimated by the marginal maximum likelihood (MML) approach, in which the latent labels are marginalized out. The expectation-maximization (EM) algorithm \citep{dempster1977maximum} is often used for this purpose, but it becomes computationally expensive and lacks theoretical guarantees in the modern regime where both $N$ and $J$ are large. 
In contrast, the joint maximum likelihood  (JML) approach adopts a fixed-effect perspective, treating the latent class labels as unknown parameters to be estimated. However, due to the inherent non-convexity of the joint likelihood function, JML is still highly sensitive to initialization and prone to be trapped in local optima. Recently,  \cite{zeng2023tensor} showed that JML is consistent under the large $N$ and large $J$ regime. Nonetheless, they do not provide any algorithmic guarantee, and the convergence rate established therein is polynomial in $J$, whose optimality remains unclear.

In this paper, we consider a different approach related to spectral clustering,  which leverages the intrinsic approximate low-rank structure of the data matrix and is computationally efficient.  In psychometrics and related fields, the common heuristic practice for spectral clustering is to first construct a similarity matrix among individuals or items, apply eigenvalue decomposition to it, and then perform K-means clustering of the eigenspace embeddings \citep{von2007tutorial, chen2017exploratory}. More recently, \citet{chen2019exploratory}  directly performs the singular value decomposition (SVD) on a regularized version of the data matrix itself. Similar ideas have also been developed for nonlinear item factor analysis \citep{zhang2020notesvd}  and the grade-of-membership model  \citep{chen2024sgom}. 

We next explain our idea of directly adapting the SVD to perform spectral clustering for binary-response LCMs. We collect the binary responses in an $N\times J$ matrix $\bR$ with binary entries, then under the typical LCM assumption, the $\bR$ can be written as 
% a low-rank signal matrix plus a mean-zero noise matrix
as $\bR=\EE[\bR]+\bE$, where $\EE[\bR]$ is the low-rank ``signal'' matrix due to the existence of $K$ latent classes, and $\bE$ is a mean-zero ``noise'' matrix. 
Denote by $\hat\bU\hat\bSigma\hat\bV^\top$ the truncated rank-$K$ SVD of $\bR$, then $\bR\approx \hat\bU\hat\bSigma\hat\bV^\top$ approximately captures the ``signal'' part. In particular, we first extract $\hat\bU$, the leading $K$ left singular vectors of the data matrix to perform dimensionality reduction. Then we apply K-means clustering algorithm on the $N$ rows of $\hat\bU\hat\bSigma$, the left singular vectors weighted by their corresponding singular values $\hat\bSigma$,
to obtain the latent class label estimates. 
While variants of this simple yet efficient algorithm have attracted significant attention in network analysis \citep{zhang2023fundamental} and sub-Gaussian mixture models \citep{zhang2022leave}, it has not been widely explored or studied in the context of LCMs. 
The only exception is the recent work \cite{lyu2025degree}, which extended spectral clustering to LCMs with individual-level degree heterogeneity and showed that spectral methods can lead to exact recovery of the latent class labels in certain regime. 
However, whether the spectral method gives an optimal way for estimating latent class labels in LCMs remains unknown.
In this paper, we resolve this question by proposing a novel two-stage method:
\begin{enumerate}
    \item \emph{First step:} Obtain  initial latent class labels  of individuals via the SVD-based spectral clustering, which leverages the approximate low-rankness of the  data matrix.
    \item \emph{Second step:} Refine the latent class labels through a single step of likelihood maximization, which ``sharpens'' the coarse initial labels using the likelihood information. 
\end{enumerate}
This hybrid approach simultaneously takes advantage of the scalability of spectral methods and leverages the Bernoulli likelihood information to achieve a vanishing exponential error rate for clustering. 
We provide rigorous theoretical guarantees for our method.
In summary, our method is both computationally fast and statistically optimal in the regime with a large number of items. {Therefore, this work provides a useful tool to aid psychometric researchers and practitioners to perform latent class analysis of modern high-dimensional data.} \color{black}{We emphasize that our method has no requirement on the magnitude of $J$ compared to $N$, and even allows $J$ to be much larger than $N$.}

The rest of this article is organized as follows. After reviewing the setup of LCMs for binary responses in \Cref{sec:model}, we discuss the performance of spectral clustering and propose our main algorithm in \Cref{sec:method}. In \Cref{sec:theory}, we establish theoretical guarantees and statistical optimality of our algorithm.
We evaluate our method's performance in numerical experiments in \Cref{sec:simulations}. The proofs of the theoretical results are included in the Appendix.

\paragraph{Notation.}
We use  bold capital letters such as $\bA,\bB,\cdots$ to denote matrices and  bold lower-case $\a,\b,\cdots$ to denote vectors. 
For any positive integer $m$, denote $[m]:=\{1,\cdots,m\}$. For any matrix $\bA\in\RR^{N\times J}$ and any $i\in[N]$ and $j\in[J]$, we use $A_{i,j}$ to denote its entry on the $i$-th row and $j$-th column, and use $\bA_{i,:}$ (or $\bA_{:,j}$) to denote its $i$-th row (or $j$-th column) vector. Let $\lambda_k\brac{\bA}$ denote the $k$-th largest singular value of $\bA$ for $k=1,\cdots,\min\ebrac{N,J}$.  Let $\op{\cdot}$ denote the spectral norm (operator norm) for matrices and $\ell_2$ norm for vectors, and $\fro{\cdot}$ denote Frobenius norm for matrices. For two non-negative sequences $\{a_N\},\{b_N\}$, we write $a_N\lesssim b_N$ (or $a_N\gtrsim b_N$) if there exists some constant $C>0$ independent of $N$ such that $a_N\le Cb_N$ (or $b_N\le Ca_N$); $a_N\asymp b_N$ if  $a_N\lesssim b_N$ and $b_N\lesssim a_N$ hold simultaneously; $a_N=o(b_N)$ if $b_N>0$ and $a_N/b_N\rightarrow 0$.

\section{Latent Class Model with Binary Responses}\label{sec:model}
% \subsection{Data Structure and Model Setup}
We consider a dataset of $N$ individuals (e.g., survey or assessment respondents, voters) and 
$J$ items  (e.g., questions in a survey or an assessment, policy items). The observed data is represented as a binary response matrix $\bR\in\ebrac{0,1}^{N\times J}$, where $R_{i,j}=1$ indicates a positive response  from individual $i$ to item $j$, and $R_{i,j}=0$ otherwise.
Each individual $i$ belongs to one of $K$ latent classes, and we encode the class membership by a latent vector $\s^\star=\brac{s_1^\star,\cdots,s_N^\star}$, where $s_i^\star\in[K]$. The interpretation of the latent classes are characterized by an item parameter matrix $\bTheta=(\theta_{j,k})\in[0,1]^{J\times K}$, where $\theta_{j,k}$
denotes the probability that an individual in class $k$ gives a positive response to item $j$.
\begin{align}\label{eq:LCM-model}
%\stackrel{}{\sim} \text{Bernoulli}\brac{
    \mathbb P(R_{i,j}=1\mid s_i^\star=k) = \theta_{j,k},\quad \forall i\in[N], j\in[J].
\end{align}
Under the LCM, a subject's $J$ responses are usually assumed to be conditionally independent given his or her latent class membership.

The traditional likelihood-based approach takes a random-effect perspective to LCMs and maximizes the marginal likelihood. This approach treats the latent class labels $\{s_i\}$ as random by assuming $ p_k=\PP\brac{s_i=k}$ for each $k$ and marginalizing $s_i$ out: 
\begin{align}\label{eq:mml-form}
    L\brac{\bTheta\mid \bR}= \sum_{i\in[N]} \log\brac{\sum_{k\in[K]}   p_k\prod_{j\in[J]} \bbrac{ \theta_{j,k}^{R_{i,j}}  
    \left(1-\theta_{j,k}\right)^{1-R_{i,j}}}}.
\end{align}
A standard approach  for maximizing  the marginal likelihood  is the EM algorithm \citep{dempster1977maximum}.
However, EM is highly sensitive to initialization. Poor initial values often lead to convergence at local maxima, so practitioners resort to multiple random restarts at substantial computational cost.   Moreover,  when  both $J$ and $K$ grow large,  the computation for evaluating the objective in \eqref{eq:mml-form}  becomes numerically unstable due to its log-sum-of-products structure.

Another strategy is to consider the fixed-effect LCM, where the joint maximum likelihood naturally comes into play. 
In particular, the  joint log-likelihood function of latent label vector $\s$ and item parameters $\mathbf\Theta$ can be written as:
\begin{align}\label{eq:llh-form}
    L\brac{\s,\mathbf \Theta\mid \mathbf R} = \sum_{i\in[N]} \sum_{j\in[J]} \bbrac{R_{i,j} \log\left(\theta_{j,s_i}\right) + 
   (1-R_{i,j}) \log\left(1-\theta_{j,s_i}\right)}.
\end{align}
The simple structure of \eqref{eq:llh-form} implies that the optimization of JML can be decoupled across $[N]$ and $[J]$. While JML is known to be inconsistent in  low-dimensional settings \citep{neyman1948consistent},
the recent study \cite{zeng2023tensor} has shown that JML can achieve consistent estimation
% and even exact recovery 
of the latent labels in the high-dimensional regime. This property makes the JML approach appealing for large-scale modern datasets.  {\color{black} It is worth noting that JML can be regarded as a version of classification EM \citep{celeux1992classification}, but with the difference of omitting the class proportion parameters. See Section 3.3 for more details.}

Throughout the paper, we consider the fixed-effect LCM and our goal is to recover the true latent labels $\s^\star$ from large-scale and high-dimensional data with $N,J\to\infty$. The performance of any clustering method is evaluated via the Hamming error up to label permutations:
\begin{align}\label{eq-hloss}
    \ell\brac{ \s, \s^\star}:=\min_{\pi\in\frakS_K}\frac{1}{N}\sum_{i\in[N]}\II\brac{ s_i\ne \pi\brac{s_i^\star}}
\end{align}
where $\frakS_K$ denotes the set of all permutation maps $\pi$ from $[K]\to[K]$. This metric accounts for the inherent symmetry in clustering structure.

\section{Two-stage Clustering Algorithm}\label{sec:method}

\subsection{Spectral clustering for the latent class model}

Our primary goal is to estimate the latent label vector $\s^\star$. However, maximizing the joint likelihood  \eqref{eq:llh-form} directly still suffers from computationally intractable for large $N$ and $J$. 
We first consider spectral clustering, which is a computationally efficient method that leverages the approximate low-rank structure of the data matrix. Let $\bZ=(Z_{i,j})\in\ebrac{0,1}^{N\times K}$ be a latent class membership matrix with $Z_{i,k}=1$ if $s^\star_i=k$ and $Z_{i,l}=0$ otherwise. Under the fixed-effect LCM,  we have $R_{i,j}\sim \text{Bern}\brac{\sum_{k=1}^{K}Z_{i,k}\theta_{j,k}}$ and we can write $R_{i,j}=\sum_{k=1}^{K}Z_{i,k}\theta_{j,k}+E_{i,j}$ where $\sum_{k=1}^{K}Z_{i,k}\theta_{j,k} = \mathbb E[R_{i,j}]$ and $E_{i,j}$ is mean-zero random noise. We have the following crucial decomposition of the data matrix  $\bR$ into two additive component using matrix notation:
\begin{align}\label{eq-decom}
 \mb R = \underbrace{\mathbf{Z}}_{N\times K} \underbrace{\mb \Theta^\top}_{K\times J} + \underbrace{\bE}_{\text{noise}}.   
\end{align}
Here, $\bZ\bTheta^\top = \mathbb E[\bR]$ is the ``signal'' part containing the crucial latent class information (and thus is low‐rank as $K\ll \min\ebrac{N,J}$), and $\bE=(E_{i,j})\in\RR^{N\times J}$ is a mean-zero ``noise'' matrix with $\mathbb E[\bE] = \mathbb E[\bR - \mathbb E[\bR]] = \bzero_{N\times J}$.

\begin{figure}[h!]
    \centering
    \includegraphics[width=0.4\textwidth]{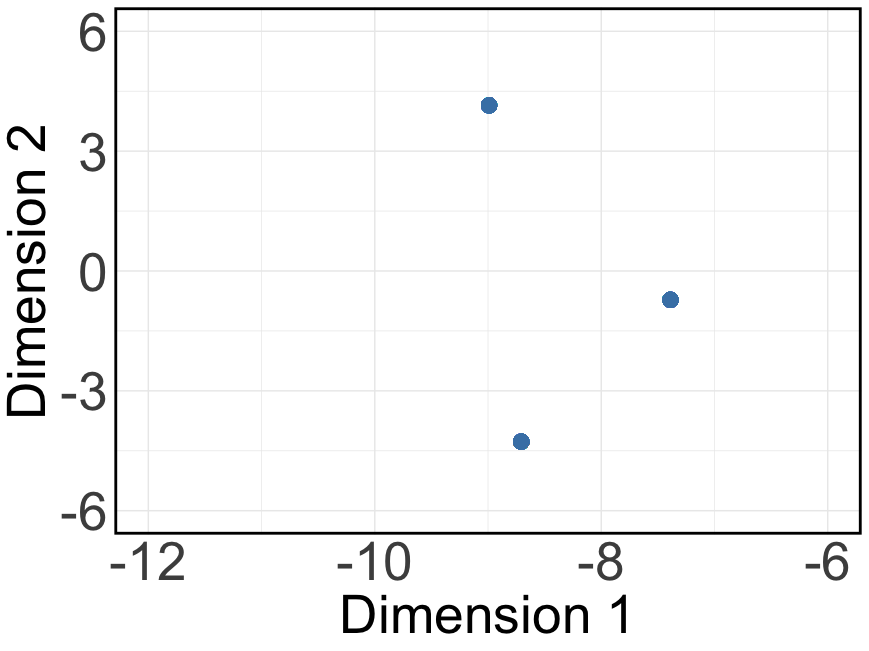}\qquad
    \includegraphics[width=0.4\textwidth]{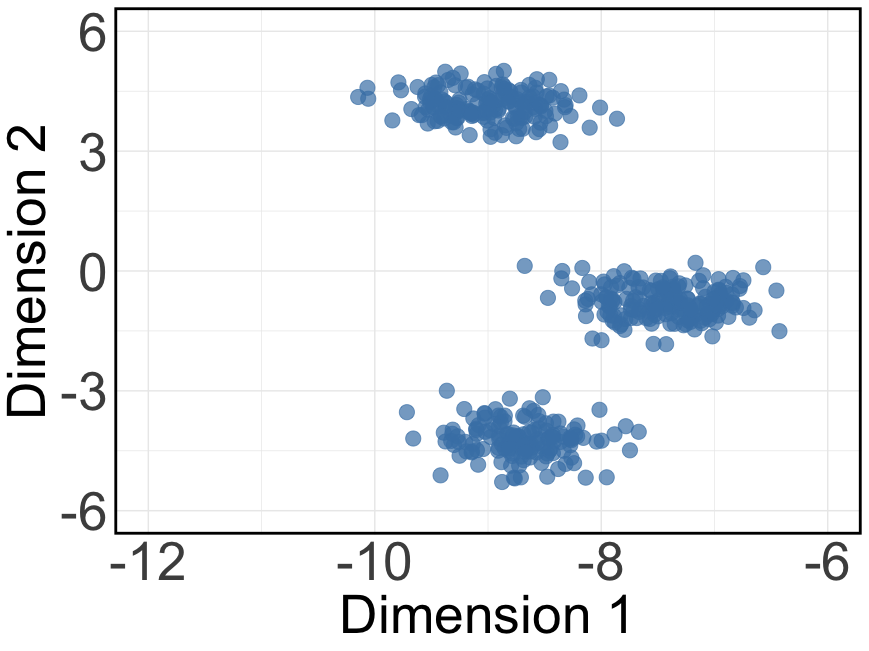}
    \caption{An illustration for spectral clustering: row vectors of $\bU\bSigma$ (left) and $\hat\bU\hat\bSigma$ (right).  Setting: $N=500$, $J=250$, $K=3$.}
    \label{fig:exmp}
\end{figure}

Given the above observation \eqref{eq-decom}, we first briefly outline the rationale of spectral clustering under the oracle noiseless setting (i.e., when $\bE=\mathbf{0}$). 
Let $\bZ\bTheta^\top = \bU\bSigma\bV^\top$ be the rank-$K$ SVD of $\bZ\bTheta^\top$. 
Notably, the $i$th row vector in $\bU\bSigma$ can be written as
\begin{align}\label{eq:exp-row-USigma}
\e_i^\top\bU\bSigma=\e_i^\top\bZ\bTheta^\top\bV=\sum_{k=1}^KZ_{i,k}\bTheta_{:,k}^\top\bV=\bTheta_{:,s_i}^\top\bV,\quad \forall i\in[N].
\end{align}
Thus, data points in the same latent class have identical row vectors in $\bU\bSigma$, namely one  of the $K$ distinct vectors  $\ebrac{\bTheta_{:,k}^\top\bV}_{k=1}^K$.
Hence the true latent class label information can be readily read off from rows of $\bU\bSigma$. In the realistic noisy data setting, denote by $\hat\bU\hat\bSigma\hat\bV^\top$ the top-$K$ SVD of the data matrix $\bR$. Then
intuitively, $\bU\bSigma$ would be well-approximated by $\hat\bU\hat\bSigma$ provided that $\bE$ is relatively small in magnitude, {and the row vectors of $\hat\bU\hat\bSigma$ are noisy point clouds distributed around the $K$ points.}
Therefore,  K-means clustering
can deliver reasonable estimates of class labels. We illustrate this point using an numerical example in \Cref{fig:exmp}, where the row vectors of $\bU\bSigma$ (left) and $\hat\bU\hat\bSigma$ (right) are plotted as points in $\mathbb R^2$, respectively. 

The above high-level idea can be formalized as the spectral clustering algorithm (\Cref{alg:Spec}), investigated in \cite{zhang2022leave} for sub-Gaussian mixture models. \Cref{alg:Spec} first obtains the rank-$K$ SVD of 
$\bR$ denoted by $\hat\bU\hat\bSigma\hat\bV^\top$, and then performs K-means clustering on the rows of $\hat\bU\hat\bSigma$, the left singular vectors weighted by the singular values.
{\color{black}
\begin{remark}
An equivalent formulation of (3) aggregates the $2^{J}$ unique response patterns (weighted by their frequencies). We do not adopt this perspective because it becomes less suitable, both computationally and theoretically, for high-dimensional data with very large $J$. On the computational side, our approach does not rely on evaluating the likelihood through these unique response patterns. On the theoretical side, our spectral method is motivated by the approximate low-rank structure of the $N\times J$ binary response matrix $\bR \in \RR^{N \times J}$ itself, which does not utilize the unique response patterns.
\end{remark}
}

\begin{algorithm}[h!]
\footnotesize
	\caption{Spectral Clustering (\textsf{Spec})}\label{alg:Spec}
	\KwData{Data matrix $\bR\in\RR^{N\times J}$. Number of latent classes $K$}
	\KwResult{Estimated latent class labels $\wt\s\in[K]^N$}
	
	Perform rank-$K$ SVD on $\bR$ to obtain $\sum_{i=1}^{K}\hat\lambda_i\hat\u_i\hat\v_i^\top $, where $\hat\lambda_1\ge \hat\lambda_2\ge \cdots \ge \hat\lambda_{K}\ge 0$,  $\hat\bU=\brac{\u_1,\cdots,\u_K}\in\RR^{N\times K}$ and $\hat\bV=\brac{\v_1,\cdots,\v_K}\in\RR^{J\times K}$.
    \\Perform $K$-means on the rows of $\hat\bU\hat\bSigma$ (or equivalently, $\bR\hat\bV$), i.e.,
\begin{align*}
        \brac{\wt\s, \ebrac{\hat\bc_k}_{k\in[K]}}=\argmin_{\s\in[K]^N,\ebrac{\bc_k}_{k\in[K]}\in \RR^{K}}\sum_{i\in[N]}\op{\hat\bU_{i,:}\hat\bSigma-\bc_{s_i}}^2.
\end{align*}
\end{algorithm}

\subsection{One-step  Likelihood Refinement with Spectral Initialization}
Despite computationally efficient, spectral clustering itself ignores the likelihood information in the latent class model that might improve statistical accuracy. This motivates our proposed \Cref{alg:SOLA} named \underline{S}pectral clustering with \underline{O}ne-step \underline{L}ikelihood refinement \underline{A}lgorithm, or SOLA for short. 
 
In particular, the first stage is to perform spectral clustering to obtain an initial class label estimates $\wt \s$. This stage utilizes the inherent approximate low-rank structure of data but does not use the likelihood information. In the second stage, recall the joint likelihood function defined in \eqref{eq:llh-form} as 
\begin{align*}
    L\brac{\s,\mathbf \Theta\mid \mathbf R} = \sum_{i\in[N]} \sum_{j\in[J]} \bbrac{R_{i,j} \log\left(\theta_{j,s_i}\right) + 
   (1-R_{i,j}) \log\left(1-\theta_{j,s_i}\right)}.
\end{align*}
We then perform the following two steps: 
\begin{enumerate}
    \item (Maximization) For fixed $\wt\s$, find the $\bTheta$ maximizing the objective  $ L\brac{\wt\s,\mathbf \Theta\mid \mathbf R}$, which turns out to have a explicit solution  as 
$$
\hat\theta_{j,k}= \frac{\sum_{i\in[N]} R_{i,j} \II\brac{\wt s_i=k}}{\sum_{i\in[N]} \II\brac{\wt s_i=k}},\quad \forall j\in[J],~ k\in[K].
$$ 
In other words, this is the sample average of $j$-th column of $\bR$ based on estimated class labels $\ebrac{i\in[N]:\wt s_i=k}$. 
    \item (Assignment) Given $\hat\bTheta$, the objective function $ L(\s,\hat \bTheta\mid \mathbf R)$ can decoupled into $N$ terms each corresponding to an individual data point, and it is equivalent to assign label $k\in[K]$ to sample $i$ which maximize the log-likelihood independently for each $i\in[N]$:
\begin{align*}
     \hat s_i = \argmax_{k\in[K]}
\sum_{j\in[J]} \Big[R_{i,j} \log\Big(\hat\theta_{j,k}\Big) + 
    (1-R_{i,j}) \log\Big(1-\hat\theta_{j,k}\Big)\Big],\quad i\in[N].
\end{align*}
\end{enumerate}
 It is readily seen that these two steps increases the joint likelihood monotonically.  However, convergence to true $\s^\star$  depends
on the performance of initial estimator $\wt \s$ in the first stage, which we obtained using spectral clustering.  The above details are summarized in \Cref{alg:SOLA} and we provide its algorithmic  guarantee  in Section~\ref{sec:theory}.

\begin{algorithm}[h!]
\footnotesize
	\caption{SOLA}\label{alg:SOLA}
	\KwData{Matrix $\bR\in\RR^{N\times J}$, rank $K$}
	\KwResult{Estimated labels $\hat\s\in[K]^N$}
	
	Initial label estimates:   $\wt \s=\textsf{Spec}\brac{\bR}$.
    \\(Maximization) Initial center estimates :
    \begin{align}\label{eq:alg-update-theta}
         \hat\theta_{j,k} = \frac{\sum_{i\in[N]} R_{i,j} \II\brac{\wt s_i=k}}{\sum_{i\in[N]} \II\brac{\wt s_i=k}},\quad  j\in[J],~ k\in[K].
    \end{align}
    \\(Assignment) One-step likelihood-based refinement:
\begin{align}\label{eq:alg-update-label}
    \hat s_i = \argmax_{k\in[K]}
\sum_{j\in[J]} \Big[R_{i,j} \log\Big(\hat\theta_{j,k}\Big) + 
    (1-R_{i,j}) \log\Big(1-\hat\theta_{j,k}\Big)\Big],\quad i\in[N].
\end{align}
\end{algorithm}

Interestingly, our two-stage procedure resonates with the spirit of the ``initialize-and-refine” scheme that has been used in network analysis for stochastic block models (SBMs) \citep{gao2017achieving,lyu2023optimal,chen2022global}, while the model considered here differs fundamentally in that LCMs handle high-dimensional response data without relying on pairwise interaction structures inherent to SBMs.

In practice,  we observe that performing multiple likelihood-based refinement steps may lead to improved convergence and performance compared to the one-step refinement procedure. Thus, we also consider a multiple-step refinement variant of  \Cref{alg:SOLA}, which we denote by \text{SOLA+}, in the later simulation studies. We set the number of ``Maximization $+$ Assignment'' refinement steps for \text{SOLA+} to be $10$ in all our simulations. {\color{black} In principle, one could also adapt the algorithm by monitoring the change in the joint log-likelihood values and stopping the algorithm once the absolute difference between successive iterations falls below a small threshold \citep{di2023two}.}

\subsection{Extension: Classification EM refinement}\label{subsec:CEM}
Our proposed framework can be readily extended to one-step Classification EM (CEM) refinement, where CEM is another widely used approach to estimate latent class models \citep{celeux1992classification, zeng2023tensor}. It can be regarded is  a variant of the standard EM algorithm designed specifically for clustering. Unlike traditional EM, which calculates ``soft" assignments by estimating the probability that each observation belongs to every latent class, CEM uses ``hard" assignments: each observation is directly assigned to the class with the highest posterior probability.  By explicitly modeling the mixture proportion parameters for the latent classes, CEM adjusts the likelihood to account for varying cluster sizes. Concretely, in our setting, one can adapt the one-step refinement in \Cref{alg:SOLA} to a CEM step by including class proportions. Besides updating \(\hat{\theta}_{j,k}\) as in \eqref{eq:alg-update-theta}, we also estimate the proportion parameters \(\hat{\p} = (\hat p_1,\ldots,\hat p_K)\in[0,1]^K\) by 
\begin{align}\label{eq:cem-update-p}
    \hat{p}_k \;=\; \frac{1}{N} \sum_{i=1}^N \mathbb{I}\{\widetilde{s}_i = k\},
    \quad k \,\in\, [K].
\end{align}
Then for each individual \(i\), we update their latent class label based on both the item parameters and the estimated class proportions:
\begin{align}\label{eq:cem-update-label}
    \hat s_i \;=\; \arg\max_{k \in [K]} \Biggl[\log \hat  p_k 
     \;+\; \sum_{j=1}^J \Bigl(\,R_{i,j}\,\log\hat{\theta}_{j,k} 
     \;+\; (1 - R_{i,j})\,\log\bigl(1 - \hat{\theta}_{j,k}\bigr)\Bigr) \Biggr].
\end{align}
This one‐step CEM update replaces the likelihood‐refinement stage of our two‐stage estimator: start with spectral clustering, then perform the CEM step above.  We provide its theoretical guarantee in Section~\ref{sec:theory}. {\color{black} In practice, one could easily generalize the above CEM procedure to a multiple-step refinement variant, which is analogous to {SOLA+}. }

\section{Theoretical Guarantees}\label{sec:theory}
To deliver the theoretical results, we start by making the following mild model assumptions.

\begin{assumption}\label{assump:dense}
   $\theta_{j,k}\in[c_\theta,C_{\theta}]$ for some universal constants $0<c_\theta<C_\theta<1$, and $j\in[J]$, $k\in[K]$.
\end{assumption}
\begin{assumption}\label{assump:balanced}
   $\min_{k\in[K]}\ab{\ebrac{i\in[N]:s_i^\star=k}}\ge \alpha N/K$ for some universal constant $\alpha\in (0,1]$.
\end{assumption}

\Cref{assump:dense} is a standard assumption for considering a compact parameter space for $\bTheta$. 
\Cref{assump:balanced} is imposed for both practical consideration and technical reason. Without this, rare classes could be statistically indistinguishable and computationally unstable to estimate.

\subsection{Theoretical Guarantee of Spectral Clustering}

In this subsection, we introduce the theoretical guarantee of spectral clustering (\Cref{alg:Spec}). To that end, we define some necessary notations. The difficulty of recovering $\bs^\star$ depends on how distinct  the latent classes are, which, in the context of spectral clustering, is characterized by  the following quantity:
\begin{align}\label{eq-def-delta}
    \Delta:=\min_{k_1\ne k_2\in[K]}\op{\bTheta_{:,k_1}-\bTheta_{:,k_2}} = \min_{k_1\ne k_2\in[K]}\left[\sum_{j=1}^J (\theta_{j,k_1} - \theta_{j,k_2})^2\right]^{1/2}.
\end{align}
This quantity measures the minimal Euclidean distance between any two class-specific item parameter vectors. Intuitively, a larger $\Delta$ yields a larger gap between the columns of $\bTheta$ and a bigger distinction among the latent classes, and therefore produces clearer separation in the spectral embeddings (referring to \eqref{eq:exp-row-USigma}), making  recovery of the true class labels easier.
In addition, let $\lambda_1\ge\cdots\ge \lambda_K>0$ denote the singular values of $\EE[\bR]=\bZ\bTheta^\top $ in decreasing order.  
The following result characterize the clustering error of \Cref{alg:Spec}.

\begin{proposition}[Adapted from Theorem 3.1 in \cite{zhang2022leave}]\label{prop:spec-clustering} Define $\sigma^2_{\theta}:=\brac{1-2\theta}/\sqbrac{2\log\brac{\brac{1-\theta}\theta^{-1}}}$ for $\theta\in(0,1)\backslash\ebrac{1/2}$ and $\sigma^2_{\theta}:=1/4$ for $\theta=1/2$, and let $\bar\sigma:=\max_{j,k}\sigma_{\theta_{j,k}}$. 
    Assume $\text{rank}\brac{\bTheta}=K$,  $\alpha N/K^2 \ge  10$ and 
    \begin{align}\label{eq:spec-cond}
    \min\ebrac{\frac{\Delta}{\alpha^{-1/2}K\brac{1+\sqrt{\frac{J}{N}}}\bar\sigma},\frac{\lambda_K}{\brac{\sqrt{N}+\sqrt{J}}\bar\sigma}}\rightarrow\infty.
    \end{align}
Then  for the spectral clustering estimator $\wt \s$ in \Cref{alg:Spec} we have
\begin{align}\label{eq-exponential}
    \EE \ell\brac{\wt \s,\s^\star}\le \exp\brac{-\frac{\Delta^2}{8\bar\sigma^2}\brac{1-o(1)}}+\exp\brac{-\frac{N}{2}},
\end{align}
where $\ell(\cdot,\cdot)$ is defined in \eqref{eq-hloss}.
Moreover, if further assume that $\Delta^2\ge 8\brac{1+c}\bar\sigma^2\log N$ for any constant $c>0$, then we have $\ell\brac{\wt \s,\s^\star}=0$ with probability $1-\exp\brac{-N/2}-\exp\brac{-\Delta/\bar\sigma}$.
\end{proposition}

\Cref{prop:spec-clustering} delivers that the simple and computationally efficient spectral clustering can essentially lead to  exponentially small error rate with respect to the latent class separation $\Delta$.  
Notably, we can have exact recovery of all latent class labels, (that is, $\ell\brac{\wt \s,\s^\star}$ with $\wt s_i=\pi(s_i)$ for all $i=1,\ldots,N$)  with high probability as long as  $\Delta^2/\bar\sigma^2\ge 8\brac{1+c}\log N$. In the LCM we study, we have $\bar\sigma\le 1/4$ by  definition and hence it suffices to have $\Delta^2\ge 2(1+c)\log N$ to achieve exact recovery.
This is not hard to achieve even for slowly growing number of items, because if we have constant separation of item parameters for each item with $\delta= \min_{k_1\neq k_2\in[K],j\in[J]} |\theta_{j,k_1} - \theta_{j,k_2}|$, then $\Delta^2 = J\delta^2$. In this case, having $J > 2(1+c)\delta^{-2} \log N$ for exact recovery corresponds to a moderately growing dimension.
In sharp contrast, the JML estimator analyzed in \cite{zeng2023tensor} only achieves a polynomial convergence rate of order $O\brac{J^{-1/2}}$. Moreover, they can not achieve exact recovery according to the technical condition and conclusion therein.  Therefore, the new spectral estimator $\wt\s$ represents a substantial improvement with exponential error decay in \eqref{eq-exponential}.

We also note that the condition  \eqref{eq:spec-cond} on $\Delta$ and $\lambda_K$ is mild and can be easily satisfied in the modern high-dimensional data setting.  We next give a concrete example. Let us consider a common generative setting for item parameters $\bTheta$ where its entries are i.i.d. generated from the Beta distribution.
\begin{proposition}\label{prop:lam-delta}
     Consider $\theta_{j,k} \stackrel{i.i.d.}{\sim} \text{Beta}(a,b)$ for all $j \in [J]$ and $k \in [K]$ with $a,b>0$. Define $$B:=\frac{a\brac{a+b+a\brac{a+b}^{-1}}}{\brac{a+b}\brac{a+b+1}},$$ 
     then under \Cref{assump:balanced} we have  $\lambda_K\gtrsim  \sqrt{\alpha B NJ}$ and $\Delta\gtrsim \sqrt{BJK}$ with probability at least $1-K\bbrac{e^{-c_1B^2J}+e^{-c_2 BJ}}$ for some universal constants $c_1,c_2\in(0,1)$.
\end{proposition}
\Cref{prop:lam-delta} entails that  the condition on $\Delta$ and $\lambda_K$ for the success of spectral clustering in \Cref{prop:spec-clustering} can be easily satisfied when $\bTheta$ consists of Beta distributed entries, a common assumption in Bayesian inference for the latent class model. Specifically, combined with \Cref{assump:dense}, a sufficient condition on $B$ for spectral clustering to have the desirable error control \eqref{eq-exponential} would be $B\gg K\brac{\frac{1}{N}+\frac{1}{J}}$, which is easily satisfied if the Beta parameters $a$ and $b$ are of the constant order.

Despite the above nice theoretical guarantee for spectral clustering, the clustering error's reliance on the Euclidean metric $\Delta$ in \Cref{prop:spec-clustering} is inherent from the general framework analyzed in \cite{zhang2022leave}, and it turns out to be not statistically optimal 
due to ignorance of the likelihood information. We will discuss and address this issue next.

\subsection{Fundamental Statistical Limit for LCMs}\label{subsec:oracle-jmle}
In this subsection, we provide the understanding of the information-theoretic limit (fundamental statistical limit) of clustering in LCMs.
To compare different statistical procedures, we consider the minimax framework that had been widely considered in both statistical decision theory and information theory in the past decades \citep{le2012asymptotic,takezawa2005introduction}. 
The minimax risk is defined as the infimum, over all possible estimators, of the maximum loss (here, the mis-clustering error) taken over the entire parameter space.  In other words, it is the rate of the best estimator one can have among all statistical procedures  in the worst-case scenario. 

Formally, we establish the  minimax lower bound for mis-clustering error under the fixed-effect LCM in the following result, whose proof is adapted from \cite{lyu2025degree} and can be found in \Cref{pf-thm:minimax-lower-bound}.
\begin{theorem}[Minimax Lower Bound]\label{thm:minimax-lower-bound} 
Consider the following  parameter space for $\EE[\bR]$:
\begin{align*}
    &\calP_K\brac{\s,\bTheta}:=\ebrac{\bar\bR:\bar R_{i,j}=\theta_{j,s_i}, \s\in[K]^{N},   \theta_{j,k}\in[0,1] , \forall i\in[N], j\in[J], k\in[K]}.
\end{align*}
Define $I^\star:=\min_{k_1\ne k_2\in[K]}\sum_{j\in[J]}I\brac{\theta_{j,k_1},\theta_{j,k_2}}$ with 
\begin{align*}
I(p,q):=-2\log\brac{\sqrt{pq}+\sqrt{\brac{1-p}\brac{1-q}}}, \quad \forall p,q\in(0,1),
\end{align*} 
then we have
    \begin{align}\label{eq:minimax-lb}
        \inf_{\hat\s }\sup_{\calP_K\brac{\s,\bTheta}}\EE\ell\brac{\hat \s,\s^\star}\ge \exp\brac{-\frac{I^\star}{2}\brac{1+o(1)}},
    \end{align}
where the infimum is taken over all possible estimators of the latent class labels.
\end{theorem}
Roughly speaking, \Cref{thm:minimax-lower-bound} informs us that under the fixed-effect LCM, the error rate $\exp\brac{-I^\star/2}$ is the lowest possible clustering error rate  that no estimator can uniformly surpass. Notably, $I(p,q)$ is exactly the Renyi divergence \citep{renyi1961measures} of order $1/2$ between two Bernoulli distributions, Bern$(p)$ and Bern$(q)$, and hence $\sum_{j\in[J]}I(\theta_{j,k_1},\theta_{j,k_2})$ can be interpreted as the Renyi divergence between two Bernoulli random vectors with parameters $\bTheta_{:,k_1}$ and $\bTheta_{:,k_2}$, repsectively. $I^\star$ can be regarded as the overall signal-to-noise ratio (SNR) which further takes account into all possible pairs of $(k_1,k_2)\in[K]^2$ with $k_1\ne k_2$.
Although the form of $I^\star$ might seem complicated at its first glance, it actually exactly reflects the likelihood information in the LCMs. By definition of the  minimax lower bound, we know that $\exp\brac{-I^\star/2}$ cannot be larger than the spectral clustering error rate $\exp\brac{-\Delta^2/(8\bar\sigma^2)}$ obtained in \Cref{prop:spec-clustering}. We will discuss this in more detail in \Cref{subsec:compare}.

In the following, we take a closer look at why the quantity $I^\star$ shows up as the fundamental statistical limit. We  analyze an idealized scenario where the true item parameters $\bTheta$ are known, which we term as the \emph{oracle} setting. For simplicity, we consider $K=2$ latent classes and $I^\star=\sum_{j\in[J]}I(\theta_{j,1},\theta_{j,2})$. In this regime, we can relate the clustering problem for the $i$-th sample to the following binary hypothesis testing problem:
\begin{align*}
    H_0:s_i^\star=1,\quad \text{v.s.} \quad H_0:s_i^\star=2.
\end{align*}
By Neyman–Pearson lemma, the test that gives the optimal Type-I plus Type-II error is the likelihood ratio test. Consequently, the oracle classifier assigns an observation to the class that maximizes the log-likelihood:
\begin{align}\label{eq:oracle-est}
    \bar s_i = \argmax_{k\in[2]}
\sum_{j\in[J]} \Big[R_{i,j} \log\theta_{j,k} + 
    (1-R_{i,j}) \log\brac{1-\theta_{j,k}}\Big].
\end{align}
Assuming without loss of generality that $s^\star_i=1$, one can show that the mis-clustering probability satisfies
\begin{align*}
    &\PP\brac{\bar s_i\ne  s^\star_i}
    =\PP\brac{\sum_{j\in[J]}R_{i,j}\log\frac{\theta_{j,2}\brac{1-\theta_{j,1}}}{\theta_{j,1}\brac{1-\theta_{j,2}}}\ge \sum_{j\in[J]}\log\frac{1-\theta_{j,1}}{1-\theta_{j,2}}}\le \exp\brac{-\frac{\sum_{j\in[J]}I(\theta_{j,1},\theta_{j,2})}{2}}.
\end{align*}
In other words, the above error rate in the oracle setting directly informs the information-theoretic limit of any clustering procedure in the LCM with binary responses, as indicated by \Cref{thm:minimax-lower-bound}.    Note that the oracle estimator \eqref{eq:oracle-est} that achieves this optimal rate is constructed using the true $\bTheta$ and cannot be applied in practice. Later we will construct a computationally feasible estimator (that is, our SOLA estimator) that achieves this optimal rate  without relying on prior knowledge of $\bTheta$.

\subsection{Theoretical Guarantee of SOLA}
In this subsection, we give the theoretical guarantee of SOLA.
To facilitate the theoretical analysis, we consider a sample-splitting variant of SOLA (\Cref{alg:SOLA-split}), where the sample data points are randomly partitioned into two subsets $\calS_1,\calS_2$ of size $\lceil N/2\rceil $ and $N-\lceil N/2\rceil $.  The key idea is to estimate $\bTheta$ using one subset of samples, and then update labels for the remaining ones. The procedure can be repeated  by switching the roles of the two subsets and we can obtain labels for all samples. 
Finally, a alignment step is needed due to the permutation ambiguity of clustering in the two subsets. 
We  have the following result, whose proof is deferred to \Cref{pf-thm:refinemnet-err-split}.

\begin{algorithm}[!ht]
\footnotesize
	\caption{SOLA with Sample-Splitting}\label{alg:SOLA-split}
	\KwData{Data matrix $\bR\in\ebrac{0,1}^{N\times J}$, index sets  $\calS_1$ and $\calS_2$,  number of clusters $K$}
	\KwResult{estimated labels $\hat \s\in[K]^{N}$}
	Construct $\bR_{\calS_1},\bR_{\calS_2}\in\ebrac{0,1}^{N/2\times J} $, which are sub-matrices of $\bR$ according to $\calS_1,\calS_2$.
	\\Initial label estimates:   $\wt \bs^{(m)}=\textsf{Spec}\brac{\bR_{\calS_m}}\in\ebrac{0,1}^{N/2}$ for $m=1,2$.
    \\Initial center estimates: for  define
		$$ \hat\theta^{(m)}_{j,k} = \frac{\sum_{i\in\calS_m} R_{i,j} \II\brac{\wt s_i^{(m)}=k}}{\sum_{i\in\calS_m} \II\brac{\wt s_i^{(m)}=k}},\quad  \forall j\in[J],~ k\in[K], ~m=1,2.$$
    \\One-step likelihood-based refinement:
\begin{align}\label{eq:jmle-label-update}
    \hat s_i ^{(m)}= \argmax_{k\in[K]}
\sum_{j\in[J]} \Big[R_{i,j} \log\Big(\hat\theta^{(m)}_{j,k}\Big) + 
    (1-R_{i,j}) \log\Big(1-\hat\theta^{(m)}_{j,k}\Big)\Big],\quad \forall i\in\calS_l,\quad l\ne m\in\ebrac{1,2}.
\end{align}
\\Label alignment:
Set $\hat s_i=\hat s_i^{(1)}$ for $\forall i\in\calS_1$. Let  $\hat\pi:=\argmin_{\pi\in\frakS_K}\fro{\hat\bTheta^{(1)}-\hat\bTheta^{(2)}\bG_{\pi}}$ and set $ \hat s_i=\hat\pi\big(\hat s_i^{(2)}\big)$ for $\forall i\in\calS_2$, where $\frakS_K$ is the set of all permutations of $[K]$ and $\bG_\pi$ is the column permutation matrix corresponding to $\pi$.
\end{algorithm}

\begin{theorem}\label{thm:refinemnet-err-split}
    Suppose Assumption \ref{assump:dense}-\ref{assump:balanced} and the assumptions of \Cref{prop:spec-clustering} hold. In addition, assume that  $N\gtrsim K\log N$, $JK\lesssim \min\ebrac{\exp\brac{-c\Delta^2/\bar\sigma^2},N^{\gamma_0}}$,
    for some universal constants $0<c<1/8$ and $\gamma_0>0$. If the signal-to-noise ratio condition satisfies    
    \begin{align*}
        \frac{\Delta/\bar\sigma}{\sqrt{J}K^{3/4}\brac{\log N}^{1/4}/N^{1/4}}\rightarrow\infty,
      \end{align*}
then for $\hat \s$ from \Cref{alg:SOLA-split} we have 
    \begin{align*}
        \EE\ell \brac{\hat \s, \s^\star}\le \exp\brac{-\frac{I^\star}{2}\brac{1-o(1)}}.
    \end{align*}
\end{theorem}
Combining \Cref{thm:minimax-lower-bound} and \Cref{thm:refinemnet-err-split}, we can see that the estimator of \Cref{alg:SOLA-split} achieves the fundamental statistical limit and is optimal for estimating the latent class labels. We have several remarks on \Cref{thm:refinemnet-err-split}. First, it is interesting to consider the regime when $J=o\brac{\sqrt{\frac{N\log N}{K}}}$, since otherwise the signal-to-noise ratio condition would already implies exact recovery (perfect estimation of latent class labels with high probability) for spectral clustering, as shown by \Cref{thm:refinemnet-err-split} and \Cref{prop:spec-clustering}. Second, the sample splitting step is crucial in the current proof of Theorem \ref{thm:refinemnet-err-split}  to decouple the dependence between $R_{i,j}$ and $\hat\bTheta$ in \eqref{eq:alg-update-label}.  
{\color{black} This step serves purely as a theoretical device to facilitate the theoretical analysis of clustering error rate, as commonly adopted in theoretical statistics \citep{lei2017generic, vu2018simple, abbe2018community, rinaldo2019bootstrapping}. In practice,  halving the sample  only increases the estimation error of $\hat{\boldsymbol{\Theta}}$ 
by a constant factor, which is negligible in theory and immaterial in practice when the sample size $N$ is large.}
Alternatively, it may be possible in the future to show that  the same theoretical guarantee holds for the original version of SOLA, i.e., \Cref{alg:SOLA}, by developing a leave-one-out analysis similar to that in  \cite{zhang2022leave}. However, the analysis will be highly technical and much more involved than that in  \cite{zhang2022leave} and beyond the scope of our current focus. Last, although sample splitting simplifies theoretical analysis, it may not be  necessary in practice; we use the original SOLA (\Cref{alg:SOLA}) for all numerical experiments and observe satisfactory empirical results.

\begin{remark}
    Our work is related to the Tensor-EM method introduced by \cite{zeng2023tensor}, which also follows an initialization-and-refinement strategy. In the high level, both our method require moment‐based methods in the  initialization.  However, there are two important distinctions. \cite{zeng2023tensor} employs a tensor decomposition that leverages higher-order moments and can be computationally intensive and unstable. In contrast,  our spectral initialization  exploits only the second moment of data (since SVD of the data matrix is equivalent to eigendecomposition of the Gram matrix), 
    making it both simple and efficient. Second, The error rate established in \cite{zeng2023tensor} is of polynomial order, i.e., $O\bigl(J^{-1/2}\bigr)$, whereas our SOLA achieves the statistically optimal exponential error rate with respect to $I^\star$.
\end{remark}

As an extension of \Cref{thm:refinemnet-err-split}, we have a  similar theoretical guarantee for the one-step CEM refinement proposed in \Cref{subsec:CEM}, whose proof is given in \Cref{sec:pf-thm:cem-refinemnet-err-split}.
\begin{corollary}\label{thm:cem-refinemnet-err-split}
    Suppose the conditions in \Cref{thm:refinemnet-err-split} hold. Consider the output of \Cref{alg:SOLA-split} with \eqref{eq:jmle-label-update} replaced by  \eqref{eq:cem-update-p} and sample splitting version of \eqref{eq:cem-update-label}, we have 
    \begin{align*}
        \EE\ell \brac{\hat \s, \s^\star}\le \exp\brac{-\frac{I^\star}{2}\brac{1-o(1)}}.
    \end{align*}
\end{corollary}
\Cref{thm:cem-refinemnet-err-split} indicates that one-step CEM refinement achieves the same exponential error rate \(\exp(-I^*/2)\) as \Cref{alg:SOLA-split}. There might be a gap between these two methods when \Cref{assump:balanced} does not hold, which is an interesting future direction to consider.

\subsection{Comparison of Spectral Clustering and SOLA}\label{subsec:compare}
In this subsection, we compare the convergence rate  of spectral clustering and SOLA under LCM. While spectral clustering is widely employed due to its computational efficiency and solid theoretical guarantees in  general clustering settings, we demonstrate that SOLA attains minimax-optimal clustering error rates in LCMs that spectral clustering fails to achieve.  

In view of \Cref{thm:minimax-lower-bound}, the minimax optimal clustering error rate for LCM is of order $\exp\brac{-I^\star/2}$. On the other hand, spectral clustering  is shown to achieve the error rate scaling as $\exp\brac{-\Delta^2/(8\bar\sigma^2)}$ in \Cref{prop:spec-clustering}. To understand the difference between these two quantities, we have the following result, whose proof can be found in \Cref{pf-prop:MLE-spectral-ratio}.
\begin{proposition}\label{prop:MLE-spectral-ratio-simple} 
\sloppy Assume that (a)  $\min_{k_1\ne k_2\in[K]}\ab{\Omega_0(k_1,k_2)}\gtrsim J$ where $\Omega_0(k_1,k_2):=\ebrac{j\in[J]:\ab{\theta_{j,k_1}-\theta_{j,k_2}}=o(1)}$, and (b) $\min_{k\in[K]}\ab{\ebrac{j\in[J]:\ab{\theta_{j,k}-1/2}>c_0}}\gtrsim J$ for some  universal constant $c_0>0$. Then we have
\begin{align*}
\exp\brac{-\frac{I^\star}{2}} &\lesssim\exp\brac{-\frac{\Delta^2}{8\bar\sigma^2}\brac{1+c}},
\end{align*}
for some universal constant $c>0$.
\end{proposition}

A few comments are in order regarding \Cref{prop:MLE-spectral-ratio-simple}.  First, under the condition (a) and (b), the rate of  spectral clustering becomes sub-optimal (in order) due to a loose constant in the exponent as the latent class separation $\Delta\rightarrow\infty$. Second, it is important to note that the mis-clustering error $\ell(\hat{\boldsymbol s}, {\boldsymbol s}^*)$ takes values in the finite set $\ebrac{0,1/N,\cdots,1}$ according to the definition \eqref{eq-hloss}.  Given this, it is only meaningful to compare error rates when the exponent $\Delta^2/\bar\sigma^2$ (or $I^\star$) does not exceed a constant factor of $\log N$, since otherwise exact clustering of all subjects is achieved  with high probability. To see this, we use Markov inequality to get  
$$
\PP(\ell(\hat{\boldsymbol s}, {\boldsymbol s}^*) \neq 0) =
\PP(\ell(\hat{\boldsymbol s}, {\boldsymbol s}^*) \ge 1/N) \leq N \EE\ell(\hat{\boldsymbol s}, {\boldsymbol s}^*) \leq N \exp\brac{-C\log N} \to 0,
$$
for a constant $C>1$.
% with $\PP\brac{\ell(\hat{\boldsymbol s}, {\boldsymbol s}^*)\ge \EE\ell(\hat{\boldsymbol s}, {\boldsymbol s}^*) }=\PP\brac{\ell(\hat{\boldsymbol s}, {\boldsymbol s}^*)\ge \exp\brac{-\log N} }=\PP\brac{\ell(\hat{\boldsymbol s}, {\boldsymbol s}^*)\ge 1/N }$ with high probability.
In view of this, the condition (a) helps us exclude the trivial case where $I^\star\asymp\Delta^2\asymp J$, which would only allow  $J\lesssim \log N$. Third, the essential order difference occurs when   most of $\theta_{j,k}$'s are bounded away from $1/2$. To see this, consider the scenario when $K=2$,   
$\theta_{j,1}=0.5$, and 
$\theta_{j,2}=0.5-\delta$ for some $\delta=o(1)$. Then we have $\Delta=\frac{1}{2}J\delta^2$ and $I^\star=-J\log\brac{\sqrt{\frac{1}{2}\brac{\frac{1}{2}-\delta}}+\sqrt{\frac{1}{2}\brac{\frac{1}{2}+\delta}}}=\frac{1}{2}J\delta^2\brac{1+o(1)}$. This implies that without condition (b), the exponents $\Delta$ and $I^\star$ are of the same order, resulting in at most a constant-factor difference in clustering error for spectral clustering and SOLA. We also conduct numerical experiments in \Cref{sec:simulations} to verify the necessity of (b).

\subsection{Estimation of the Number of Latent Classes $K$}
In practice, the true number of latent classes $K$ may be unknown.  To see how we can estimate $K$ from data, recall that $\bR=\bZ\bTheta^\top+\bE$. By Weyl's inequality and classical random‐matrix theory \citep{vershynin2018high}, the singular values of $\bR$ satisfy $$\ab{\lambda_i\brac{\bR}-\lambda_i\brac{\bZ\bTheta^\top}}\le \op{\bE}=O_p\brac{\sqrt{J}+\sqrt{N}},\quad i\in[N\wedge J]$$ 
Since $\bZ\bTheta^\top$ has exactly $K$ nonzero singular values $\lambda_1\ge \cdots\ge \lambda_K>0$ and  \Cref{prop:lam-delta} shows that $\lambda_K\gtrsim  \sqrt{\alpha BNJ}$ under a common generative setting, the top $K$ singular values of $\bR$  lie well above the noise level. Accordingly, we estimate $K$ by counting how many singular values of 
$\bR$ exceed a threshold slightly above the noise level:
\begin{align}\label{eq:K-hat}
    \hat K=\ab{\ebrac{i\in[N\wedge J]:\lambda_i\brac{\bR}>2.01 \brac{\sqrt{J}+\sqrt{N}}}}.
\end{align}
Here the factor $2.01$  ensures we count only those singular values contributed by the low-rank signal.  We have the following lemma certifying the consistency of $\hat K$, whose proof is deferred to \Cref{sec:proof-K-est}.
\begin{lemma}\label{lem:K-est}
    Instate the assumptions of \Cref{prop:spec-clustering}. For $\hat K$ defined in \eqref{eq:K-hat}, we have $$\PP\brac{\hat K=K}=1-o(1),\qquad \text{as~}N,J\rightarrow\infty.$$
\end{lemma}
This thresholding rule provides a formal analogue of the classical scree‐plot method \citep{cattell1966scree, zhang2020notesvd}, and its proof shares a similar  spirit of Theorem 2 in \cite{zhang2020notesvd} but with a sharper characterization of the constant factor in the noise magnitude.

\section{Numerical Experiments}\label{sec:simulations}
\subsection{Simulation Studies}
We compare our proposed methods SOLA (\Cref{alg:SOLA}) and  $\text{SOLA}+$ (\Cref{alg:SOLA} with multiple steps refinement), with spectral method \citep{zhang2022leave}, EM \citep{linzer2011polca}, and Tensor-CEM \citep{zeng2023tensor}. For the EM algorithm, we employ the popular \texttt{polca} package \citep{linzer2011polca} in R, and for Tensor-CEM, we use the original Matlab code \citep{zeng2023tensor} for tensor initialization combined with our R-based implementation of the CEM refinement. Our evaluation focuses on three aspects: (a) clustering accuracy, (b) stability, and (c) computational efficiency. Throughout the simulations, we generate the true latent label   $\ebrac{s_i^\star}_{i=1}^N$  independently uniformly  from $[K]$ and  set $K=3$, $N=2J$.  The parameters $\theta_{j,k}$
are independently sampled from Beta$(\beta_1,\beta_2)$. We generate $200$ independent replicates and report the average mis-clustering error and computation time, excluding any replicates in which the algorithm fails (Tensor-CEM and EM). Notably, our algorithm is numerically stable  without any failures in all simulations. 

\color{black}
\begin{remark}[EM implementations: \texttt{polca} vs.\ \textsc{LatentGOLD}]
Beyond \texttt{polca}, we also evaluated \textsc{LatentGOLD} \citep{vermunt2005latent}, which implements a robust EM routine with safeguards against numerical underflow. {\Cref{tab:Comparison-latentgold} reports a representative comparison across three regimes ($N=2J$, $N=J$, and $N=J/2$), showing that \textsc{LatentGOLD} is substantially more stable and accurate than \texttt{polca} in these settings.
Our overall message remains unchanged:  EM-type methods can be competitive in the classical regime with relatively large $N$ and smaller $J$, whereas SOLA/SOLA+ become increasingly advantageous as $J$ grows and the problem enters the high-dimensional regime (with $J$ comparable to or larger than $N$).  While retaining \texttt{polca} for comparison purposes in our large-scale simulation studies, we emphasize that when analyzing data with a small-to-moderate $J$, it is recommended to use \textsc{LatentGOLD} for their better numerical stability  and performance.}
\end{remark}

\begin{table}[!htbp]
\centering
\begin{tabular}{|c|c|c|c|c|}
\hline
Setting \textbackslash~ Method           & SOLA  & SOLA+  & EM by \texttt{polca} & EM by \textsc{LatentGold} \\ \hline
$N=190, J=95$ ($N=2J$)   & 0.0815 &  0.0772 &  0.3022 & 0.0645      \\ \hline
$N=95, J=95$ ($N=J$) & 0.1760 & 0.1745 &  0.4642 & 0.2609 \\ 
\hline
$N=75, J=150$ ($N=J/2$) & 0.0411 & 0.0411 &  0.5002 & 0.2795 \\ 
\hline
\end{tabular}
 \caption{Clustering error across different methods and settings under $100$ replicates } 
    \label{tab:Comparison-latentgold}
\end{table}
\color{black}

\begin{figure}[h!]
    \centering
    \includegraphics[width=0.5\textwidth]{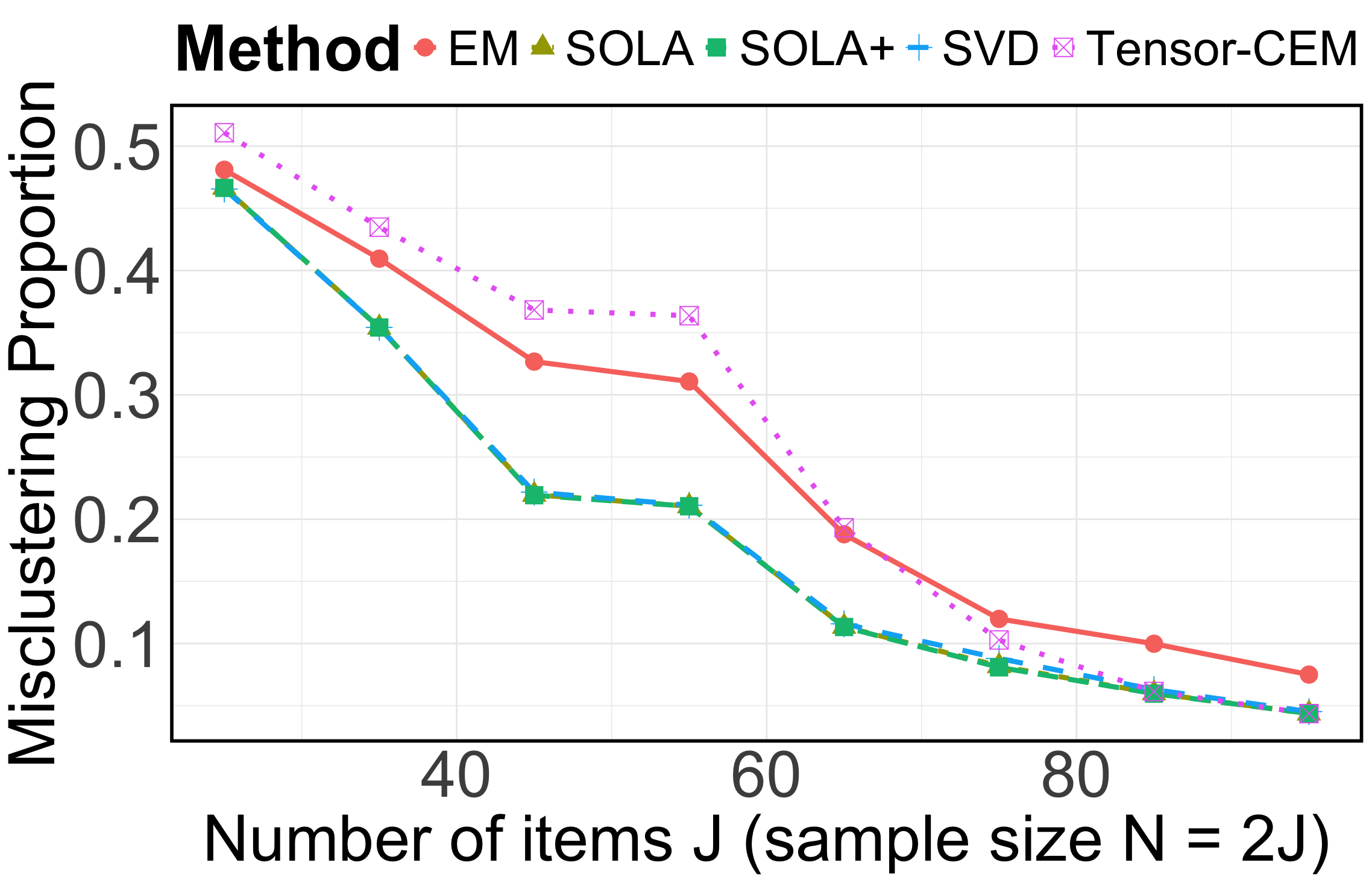}
    \caption{Simulation 1: Mis-clustering proportions v.s. number of items $J$.  Entries of $\bTheta$ are independently  generated from  $\text{Beta}\brac{5,5}$.}
    \label{fig:sim-1}
\end{figure}

\paragraph{Simulation 1: Latent class estimation accuracy for dense $\bTheta$ close to $1/2$.}
In the first simulation, we set $(\beta_1,\beta_2)=(5,5)$. Under these parameters, the entries of $\bTheta$ tend to be close to $0.5$. As shown in Figure~\ref{fig:sim-1}, our proposed methods achieve lower mis-clustering errors compared to the alternative approaches. In this setting, the improvement of SOLA over the plain spectral clustering method is negligible, which is expected since $\theta_{j,k}$'s values near $0.5$ reduce the additional gain from incorporating the likelihood information. This observation supports the necessity of having $\theta_{j,k}$'s  well separated from $0.5$ (see Proposition~\ref{prop:MLE-spectral-ratio}) to have likelihood-based refinement improve upon spectral clustering. It is also worth noting that Tensor-CEM performs poorly when both the number of items $J$ and the sample size $N$ are small, likely due to instability introduced by its tensor-based initialization step.

\begin{figure}[h!]
    \centering
    \includegraphics[width=0.5\textwidth]{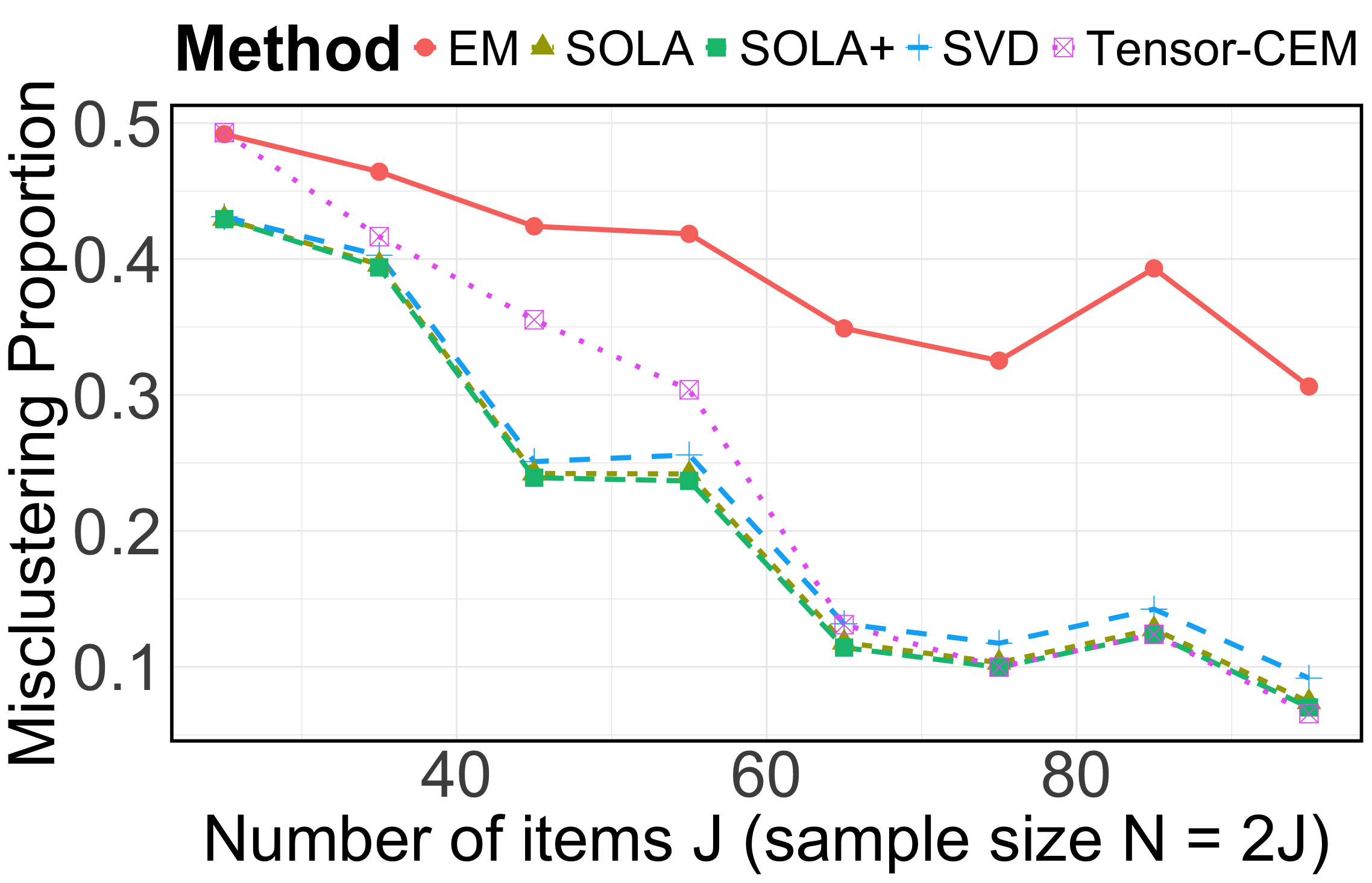}
    \caption{Simulation 2: Mis-clustering proportions v.s. number of items $J$.  Entries of $\bTheta$ are independently  generated from $\text{Beta}\brac{1,8}$.}
    \label{fig:sim-2}
\end{figure}

\paragraph{Simulation 2: Latent class estimation accuracy for sparse item parameters $\bTheta$.}
In the second simulation, we choose $(\beta_1,\beta_2)=(1,8)$, a sparse scenario where the $\theta_{j,k}$'s tend to be close to $0$. Figure~\ref{fig:sim-2} illustrates that our proposed methods exhibit superior clustering accuracy when $N$ and $J$ are small, while maintaining competitive performance as these dimensions increase. In this sparse setting, the traditional EM algorithm performs the worst because $\theta_{j,k}$'s are near the boundary (close to $0$), making it difficult for random initializations to converge to optimal solutions. Furthermore, the gap between spectral clustering and SOLA becomes more pronounced when the $\theta_{j,k}$'s  are bounded away from $0.5$.

\begin{figure}[h!]
    \centering
    \includegraphics[width=0.5\textwidth]{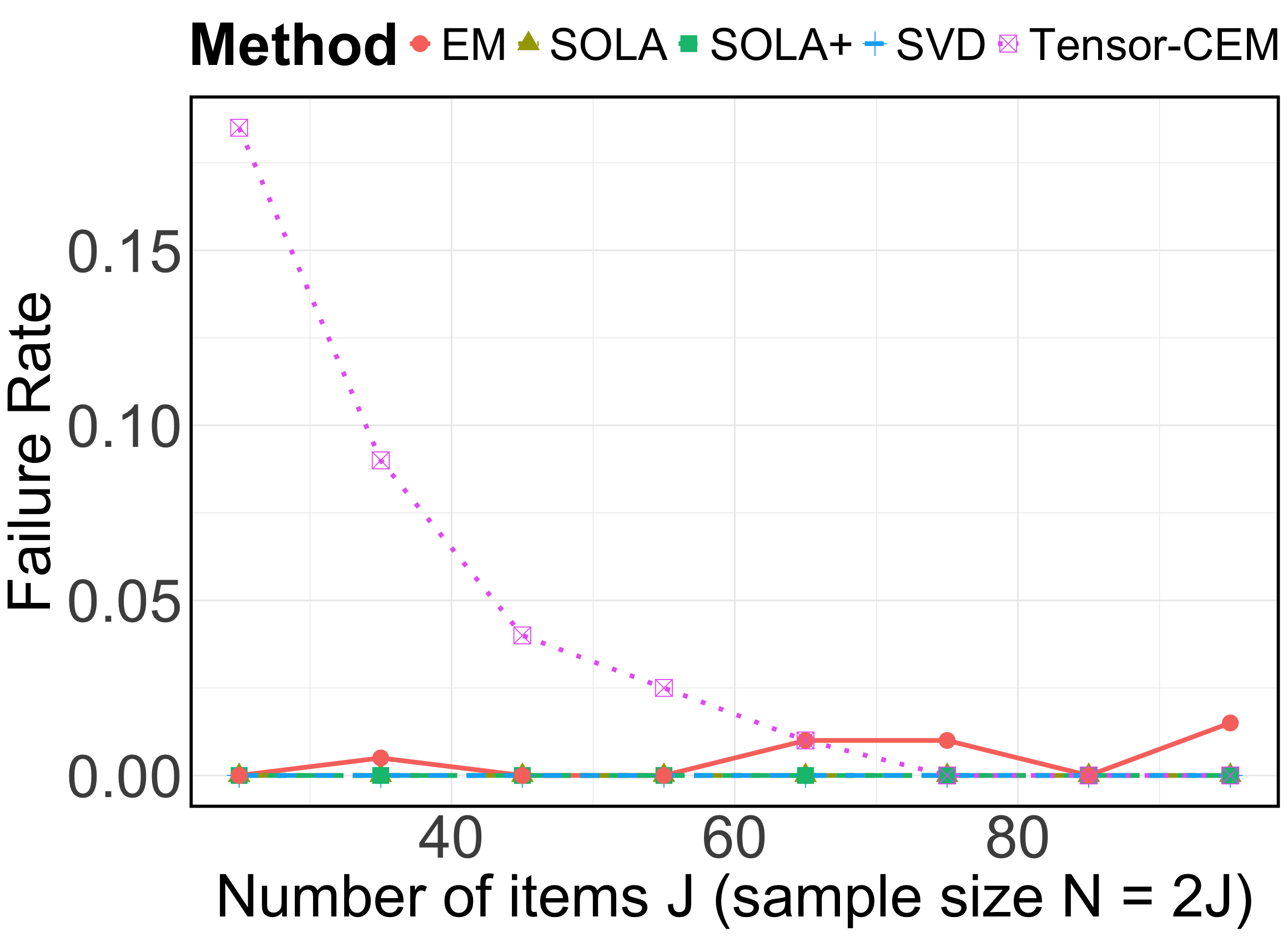}
\hfill
    \caption{Simulation 3: Failure rate versus the number of items $J$.  Entries of $\bTheta$ are independently  generated from $\text{Beta}\brac{1,8}$. }
    \label{fig:sim-3}
\end{figure}

\paragraph{Simulation 3: Stability analysis.}
We next assess the stability of the different methods under the sparse setting $(\beta_1,\beta_2)=(1,8)$. In particular, we record a ``failure" point if either (a) the estimated number of classes degenerates from the pre-specified $K$ to a smaller value, or (b) the algorithm produces an error during execution. As shown in Figure~\ref{fig:sim-3}, our proposed methods are robust across a wide range of $N$ and $J$ without any failures in any simulation trial. In particular, the failure rate for Tensor-CEM exceeds 15\% for small $N$ and $J$, likely due to its sensitive initialization, while the EM algorithm becomes increasingly unstable as the dimensions grow. These results underscore the robustness of our SOLA approaches.

\begin{figure}[h!]
    \centering
    \includegraphics[width=0.5\textwidth]{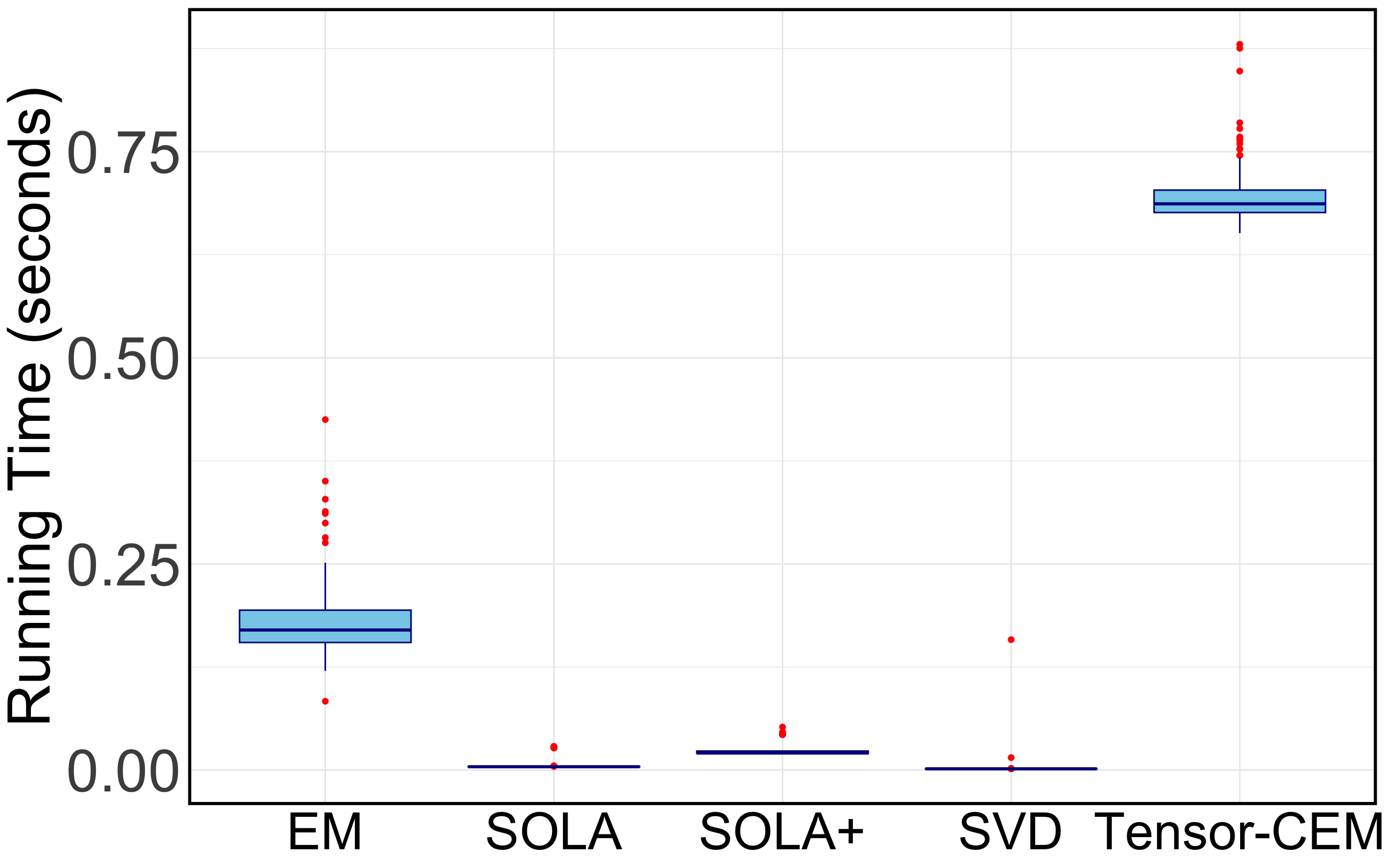}
\hfill
    \caption{Simulation 4: Running time (seconds) of different methods. }
    \label{fig:sim-4}
\end{figure}

\paragraph{Simulation 4: Computational efficiency.}
Finally, we compare the computation time required by each method. In our experiments, the SVD step and our SOLA and  $\text{SOLA}+$ algorithms are implemented in R. The total computation time for Tensor-CEM is the sum of the Matlab initialization and the R refinement stages. As reported in Table~\ref{table:sim-4} and illustrated in Figure~\ref{fig:sim-4}, our proposed methods are significantly faster than both the EM and Tensor-CEM approaches, with computation time comparable to that of  SVD.

\begin{table}[h!]
\centering
\begin{tabular}{|c|c|c|c|c|c|}
\hline
     & SVD               & SOLA    & $\text{SOLA}+$             & EM               & Tensor-CEM        \\
     \hline
$J=25$ & $4.15\times 10^{-4}$ & $8.62\times 10^{-4}$ & $4.19\times 10^{-3}$ & $1.42\times 10^{-2}$ & $5.12\times 10^{-1}$ \\ \hline
$J=95$ & $2.33\times 10^{-3}$ & $4.66\times 10^{-3}$ & $2.51\times 10^{-2}$ & $1.80\times 10^{-1}$ & $6.95\times 10^{-1}$ \\ \hline
\end{tabular}
\caption{Simulation 4: Running time (seconds) of different methods. }
\label{table:sim-4}
\end{table}

\begin{figure}[h!]
    \centering
    \includegraphics[width=0.5\textwidth]{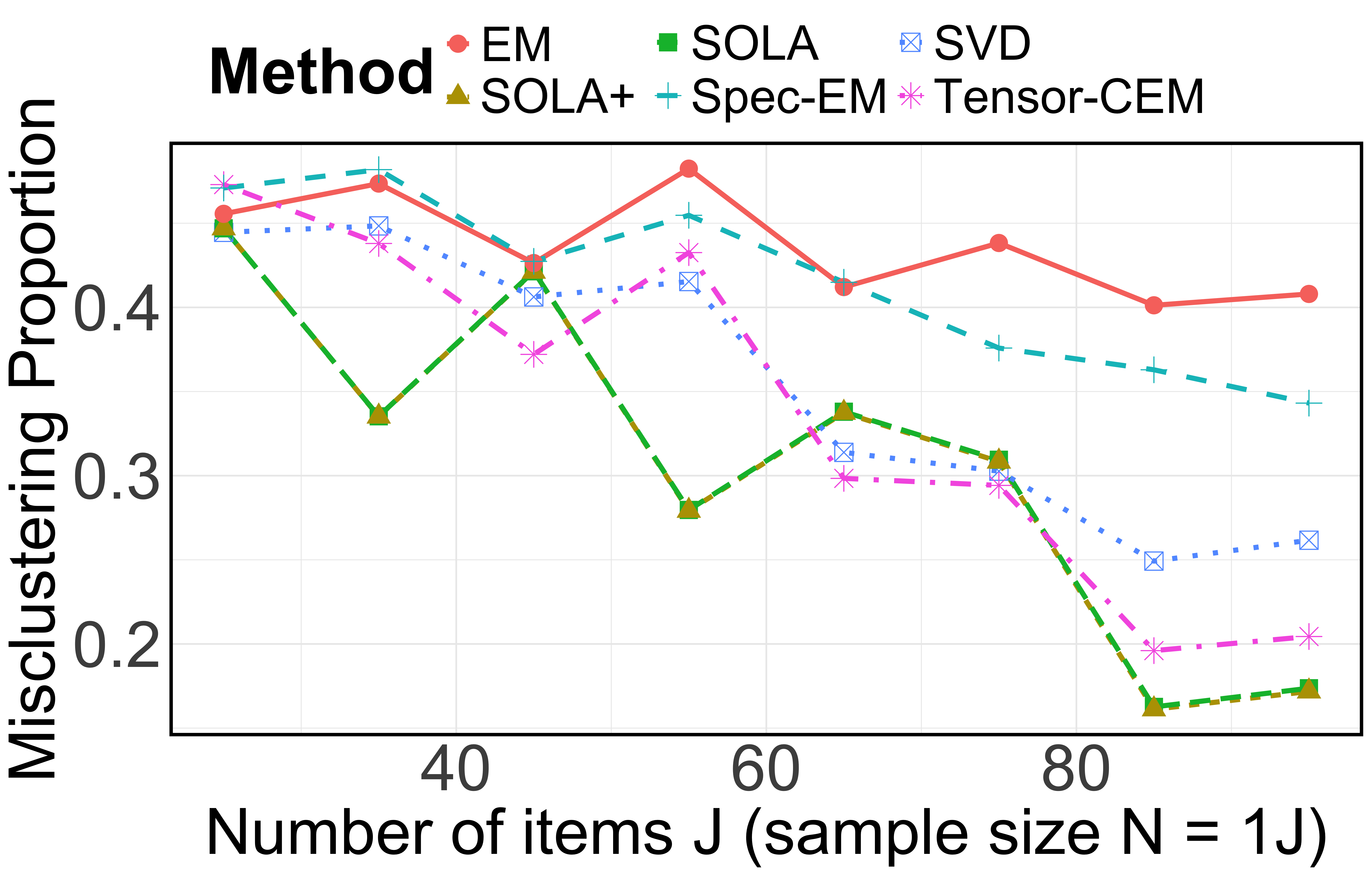}
\hfill
    \caption{\color{black}Simulation 5-1: Mis-clustering proportions v.s. number of items $J$ under $N=J$.  Entries of $\bTheta$ are independently  generated from $\text{Beta}\brac{1,8}$.}
    \label{fig:sim-N=J}
\end{figure}

\begin{figure}[h!]
    \centering
    \includegraphics[width=0.5\textwidth]{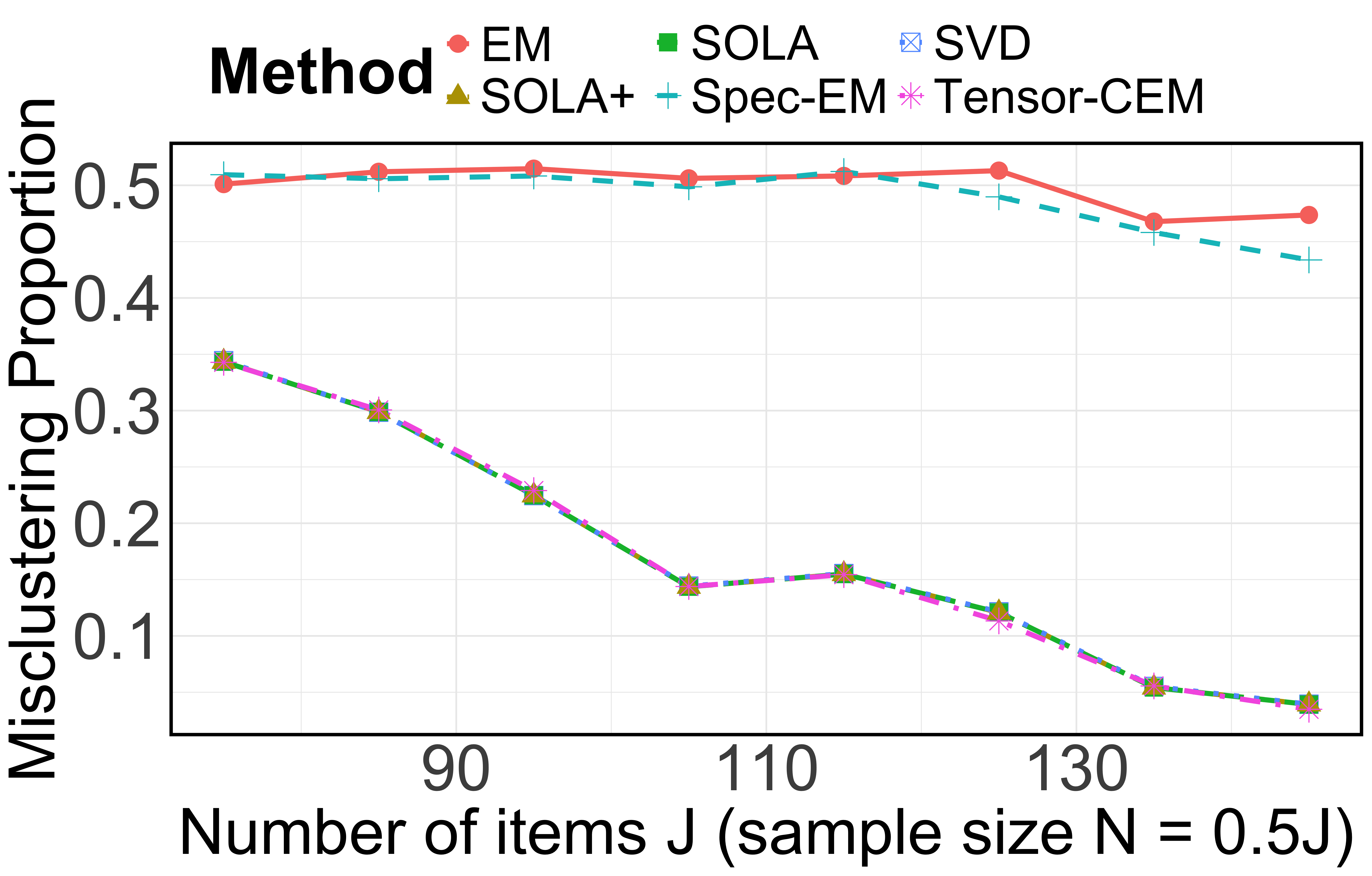}
\hfill
    \caption{\color{black}Simulation 5-2: Mis-clustering proportions v.s. number of items $J$ under $N=0.5J$.  Entries of $\bTheta$ are independently  generated from $\text{Beta}\brac{1,8}$.}
    \label{fig:sim-N=0.5J}
\end{figure}

\begin{figure}[h!]
    \centering
    \includegraphics[width=0.5\textwidth]{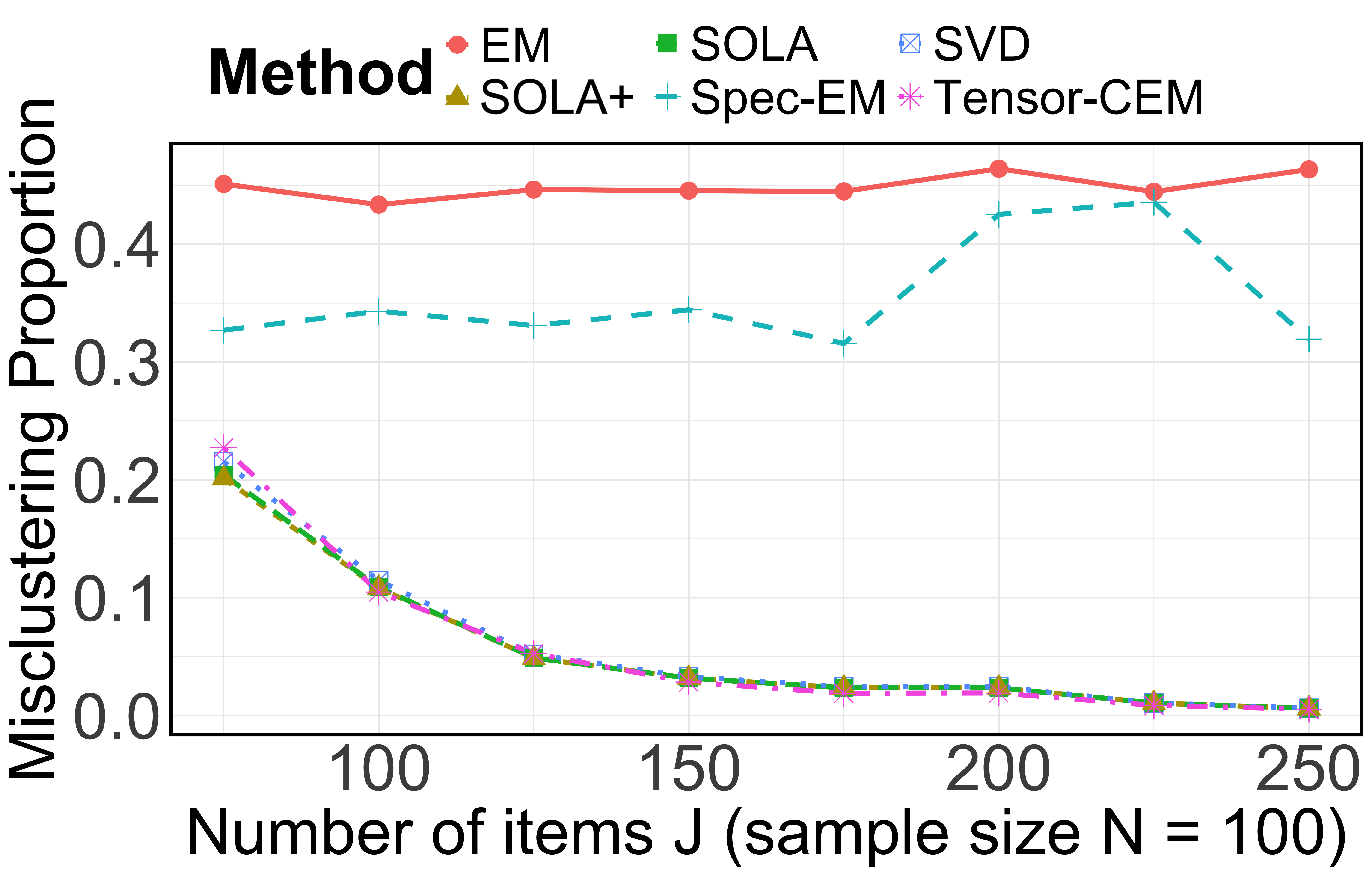}
\hfill
    \caption{\color{black}Simulation 5-3: Mis-clustering proportions v.s. number of items $J$ under $N=100$.  Entries of $\bTheta$ are independently  generated from $\text{Beta}\brac{1,8}$.}
    \label{fig:sim-N=100}
\end{figure}

\begin{figure}[h!]
    \centering
    \includegraphics[width=0.5\textwidth]{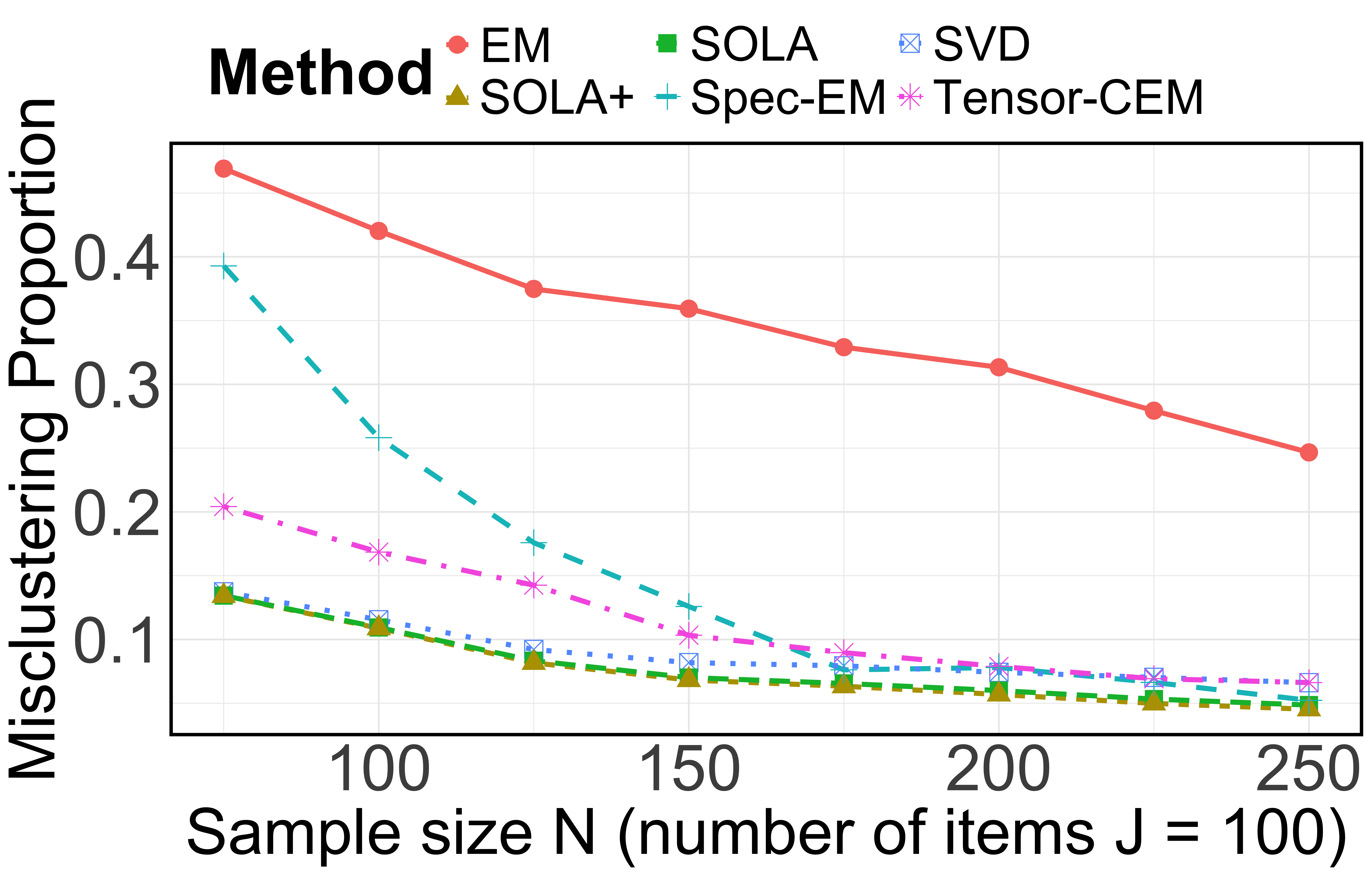}
\hfill
    \caption{\color{black}Simulation 5-4: Mis-clustering proportions v.s. sample size $N$ under $J=100$.  Entries of $\bTheta$ are independently  generated from $\text{Beta}\brac{1,8}$.}
    \label{fig:sim-J=100}
\end{figure}

\begin{figure}[h!]
    \centering
    \includegraphics[width=0.5\textwidth]{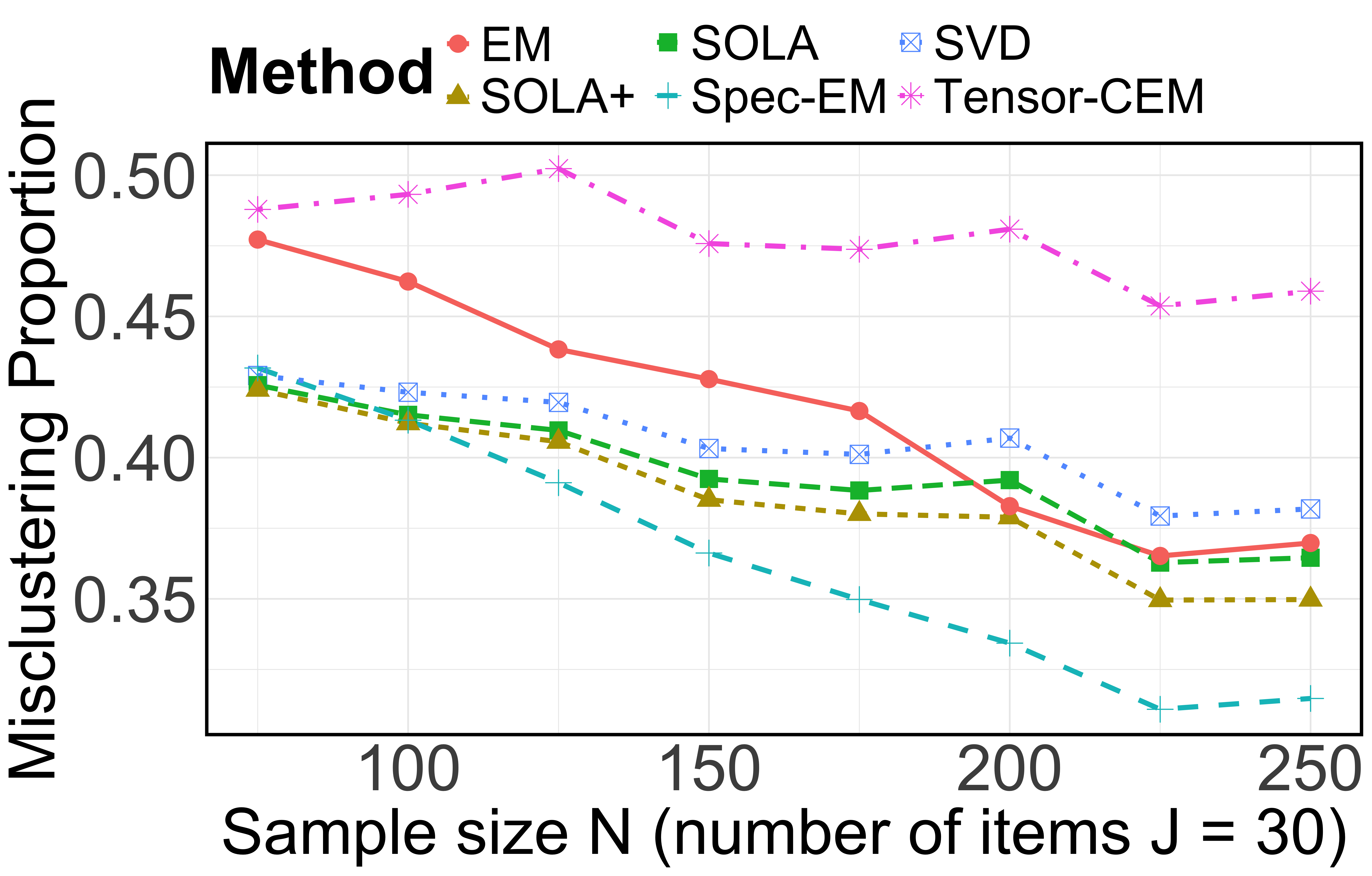}
\hfill
    \caption{\color{black}Simulation 5-5: Mis-clustering proportions v.s. sample size $N$ under $J=30$.  Entries of $\bTheta$ are independently  generated from $\text{Beta}\brac{1,8}$.}
    \label{fig:sim-J=30}
\end{figure}

\begin{figure}[h!]
    \centering
    \includegraphics[width=0.5\textwidth]{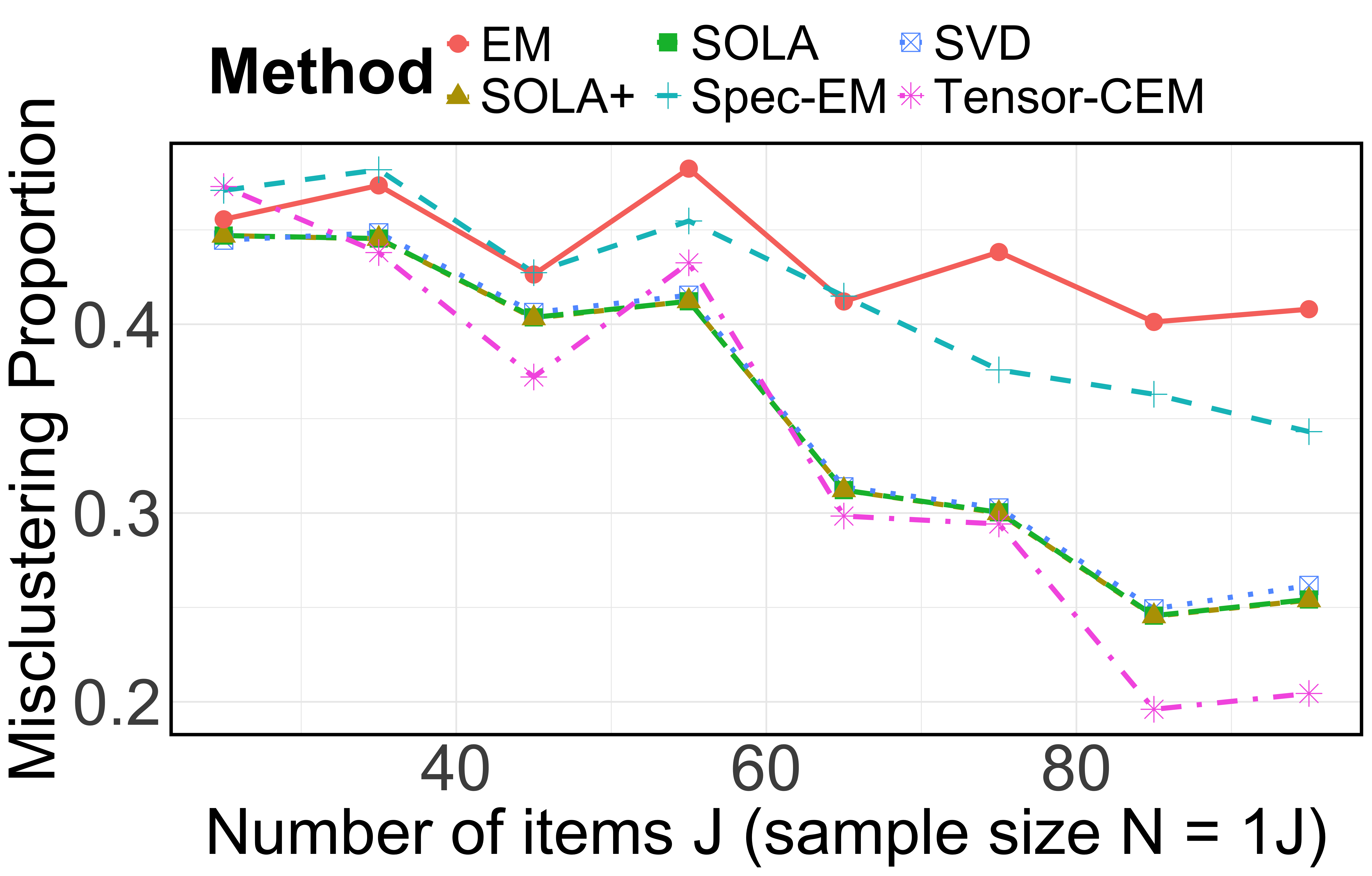}
\hfill
    \caption{\color{black}Simulation 5-6: Mis-clustering proportions v.s. number of items $J$ under imbalanced latent classes with $N=J$. Entries of $\bTheta$ are independently  generated from $\text{Beta}\brac{1,8}$.}
    \label{fig:sim-imbalanced}
\end{figure}

\begin{figure}[h!]
    \centering
    \includegraphics[width=0.5\textwidth]{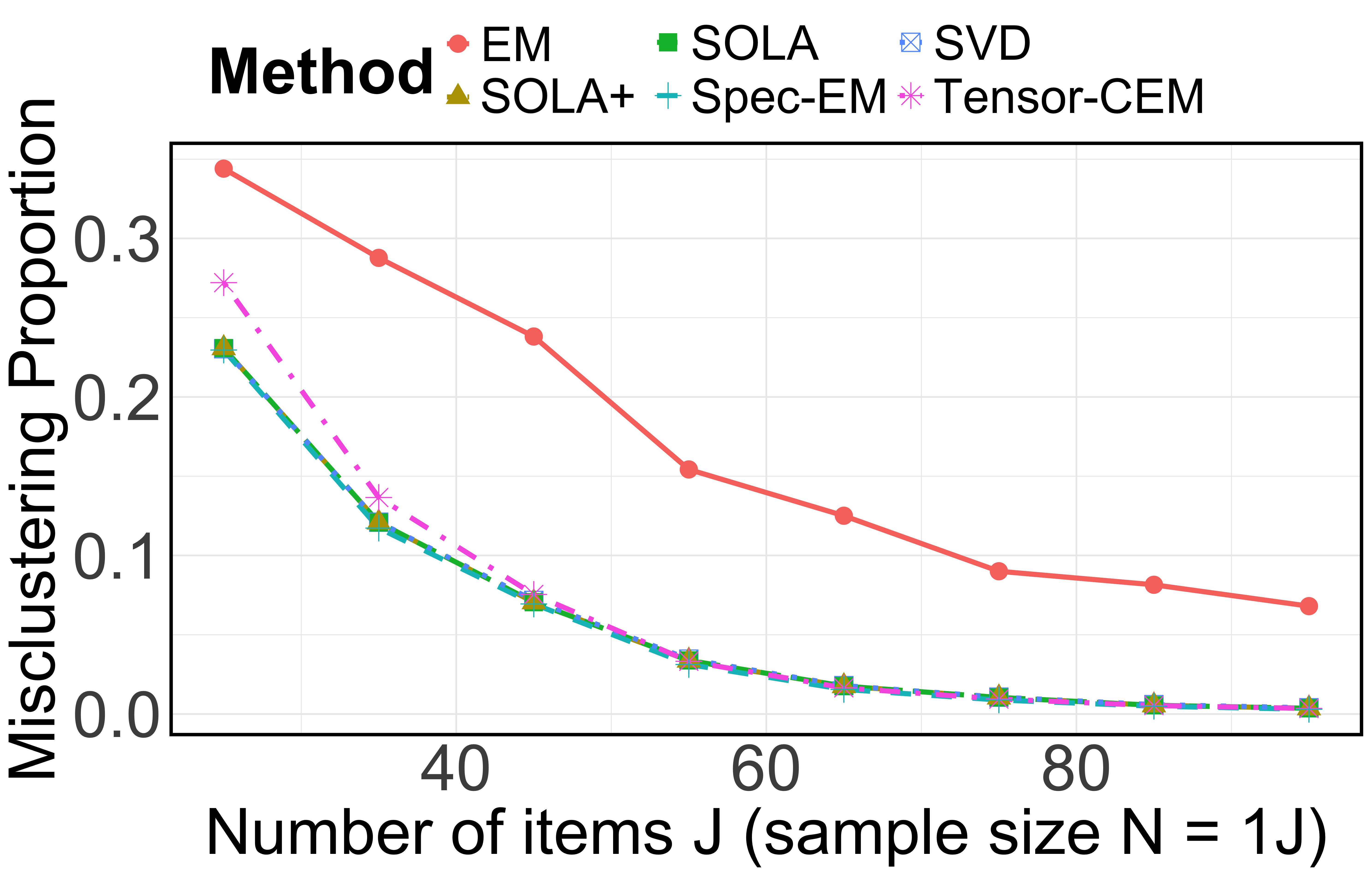}
\hfill
    \caption{\color{black}Simulation 5-7: Mis-clustering proportions v.s. number of items $J$ under $N=J$. The $\bTheta$ matrix is set to a fixed design matrix.}
    \label{fig:sim-fixed-design}
\end{figure}

\color{black}
\paragraph{Simulation 5: Latent class estimation accuracy under different regimes.}
In the fifth simulation study, we also employ Spec-EM (EM algorithm initialized by spectral clustering), motivated by the fact that when the number of items is moderately large, spectral clustering  provides an effective initialization for the classical EM algorithm for LCMs.
In particular, we consider the following regimes: (i) $N=J$ (\Cref{fig:sim-N=J}), (ii) $N=J/2$ (\Cref{fig:sim-N=0.5J}), (iii) fix $N=100$ and vary $J$ (\Cref{fig:sim-N=100}), (iv) fix $J=100$ and vary $N$ (\Cref{fig:sim-J=100}), (v) fix $J=30$ and vary $N$ (\Cref{fig:sim-J=30}), (vi) $N=J$ with $\ebrac{s_i^\star}_{i=1}^N$ being independently generated from $[3]$ with probability $(p_1,p_2,p_3)=(1/6,1/3,1/2)$ (\Cref{fig:sim-imbalanced}). In all these settings, we set $(\beta_1,\beta_2)=(1,8)$. In addition, we consider (vii) $N=J$ under the fixed design of $\bTheta$ (\Cref{fig:sim-fixed-design}), where the fixed item parameter matrix $\boldsymbol\Theta$ is constructed by vertically concatenating copies of a $5\times 3$ block matrix $\boldsymbol\Theta_0$ as follows:
\[
\boldsymbol\Theta=\begin{bmatrix}
  \boldsymbol\Theta_0\\
  \vdots\\
   \boldsymbol\Theta_0
\end{bmatrix}\Bigg\}~b~\text{copies},\qquad \boldsymbol\Theta_0:=\begin{bmatrix}
0.3 & 0.7 & 0.7\\
0.3 & 0.7 & 0.3\\
0.7 & 0.3 & 0.7\\
0.7 & 0.3 & 0.3\\
0.7 & 0.7 & 0.3
\end{bmatrix}.
\]
Here the block matrix $\boldsymbol\Theta_0$ is vertically stacked $b$ times to yield $J=5b$ items with $K=3$ latent classes. It is clear that { SOLA} and { SOLA+} perform significantly better in most scenarios except the most traditional large-$N$, small-$J$ regime, where Spec-EM performs better. This underpins both the validity of SOLA and spectral clustering.

\color{black}
\paragraph{Simulation 6: Estimation of number of latent classes.}
Our theoretical guarantee for the selection criterion of $K$ relies on random matrix theory under the asymptotic regime $N,J \to \infty$. This means relatively large $N$ and $J$ are needed for the theoretical guarantee to hold, which is consistent with our focus on high-dimensional data. The constant ``2" in the threshold reflects a worst-case bound and can indeed be conservative, especially when entries of $\boldsymbol{\Theta}$ are mostly concentrated away from $1/2$. To illustrate its effectiveness in practice, we have performed  simulation studies with $N=2J$ and $K=3$, under several generative mechanisms for $\boldsymbol{\Theta}$ (i.i.d. Beta$(1,1)$, Beta$(1,4)$, Beta$(1,5)$, which reflect different sparsity levels, and also the fixed design as described in Simulation 5). As shown in \Cref{tab:Comparison}, the criterion successfully recovers $K$ in high-dimensional regimes. Our real-data analysis in Section 5.2 also demonstrates that the method works well when the data are sufficiently large. For smaller $N$ and $J$, we recommend that practitioners use classical information criteria, such as BIC, as a more reliable and interpretable tool for selecting $K$.

\begin{table}[h!]
\def\arraystretch{1.4}%
\centering
\begin{tabular}{c|c|c|c|c|c|c}
\hline
 $J$   & $ 1000$ & $1200$ & $1400$ & $1600$ & $1800$ & $2000$\\ \hline
Fixed design of item parameters &   $100\%$   &  $100\%$    &  $100\%$  &  $100\%$  &  $100\%$  &  $100\%$   \\ \hline
$\theta_{j,k}\stackrel{i.i.d.}{\sim}$Beta$(1,1)$ &   $100\%$   &  $100\%$    &  $100\%$  &  $100\%$  &  $100\%$  &  $100\%$                           \\ \hline
$\theta_{j,k}\stackrel{i.i.d.}{\sim}$Beta$(1,4)$ &   $0\%$   &  $0\%$    &  $100\%$  &  $100\%$  &  $100\%$  &  $100\%$                            \\ \hline
$\theta_{j,k}\stackrel{i.i.d.}{\sim}$Beta$(1,5)$ &   $0\%$   &  $0\%$    &  $0\%$  &  $0\%$  &  $70\%$  &  $100\%$                            \\ \hline
\end{tabular}
 \caption{\color{black}Proportion of successfully selecting $\hat K$ to be the true $K$ based on $200$ simulation replicates, under different generative mechanisms of the item parameters $\boldsymbol\Theta=(\theta_{j,k})_{J\times K}$.} 
    \label{tab:Comparison}
\end{table}

\color{black}
\subsection{Real Data Example}\label{subsec:real-data}
We further evaluate the performance of our methods on the United States 112th Senate Roll Call Votes data, which is publicly available at \url{https://legacy.voteview.com/senate112.htm}. The dataset contains voting records for $J=486$ roll calls by $102$ U.S. senators. Following the preprocessing procedure in \cite{lyu2025degree}, we convert the response matrix to binary and remove senators who are neither Democrats nor Republicans, as well as those with more than $10\%$ missing votes. This results in a final sample of $N=94$ senators. For the remaining missing entries, we impute values by randomly assigning $0$ or $1$ with probability equal to the positive response rate observed from each senator's non-missing votes.

We compare the performance of the methods considered in our simulation studies. Table~\ref{table:real} summarizes the mis-clustering error and computation time (in seconds) for each method. Interestingly, the SVD approach yields an error comparable to those obtained with \textsf{SOLA} and \textsf{SOLA+}, suggesting that the refinement step may be unnecessary for this dataset. This observation can be partially explained by our remark following Theorem~\ref{thm:refinemnet-err-split}: when the number of items $J$ is substantially larger than the number of individuals $N$, spectral clustering alone can perform very well. In contrast, the traditional EM method exhibits significantly higher mis-clustering error, and EM and Tensor-CEM suffers from high computational cost in this large-$J$ regime.

\begin{table}[h!]
\centering
\begin{tabular}{|c|c|c|c|c|c|}
\hline
     & SVD               & SOLA    & $\text{SOLA}+$             & EM               & Tensor-CEM        \\ \hline
Error & $0.0212$ & $0.0212$ & $0.0212$ & $0.117$ & $0.0212$ \\ \hline
Time & $0.00179$ & $0.0344$ & $0.0503$ & $3.125$ & $5.154$ \\ \hline
\end{tabular}
\caption{Mis-clustering error and running time (seconds) of different methods on Senate Roll Call Votes data. }
\label{table:real}
\end{table}

We also validate the performance of our method for estimating $K$ on this real dataset. In particular, the threshold defined in \eqref{eq:K-hat} is $2.01\times (\sqrt{486}+\sqrt{94})\approx63.8$. The first three largest singular values of $\bR$ are $\{148.1,\,64.4,\,16.6\}$, which leads to an estimated number of clusters $\hat K=2$. Notably, this estimated value exactly matches the known number of clusters in the dataset (i.e., \(K=2\)).

\section{Discussion}
In this work, we have proposed SOLA, a simple yet powerful two‐stage algorithm, under binary-response latent class models. Our method efficiently exploits the low-rank structure of the response matrix and further leverage the likelihood information. We have proved that SOLA attains the optimal mis‐clustering rate and achieves the fundamental statistical limit. We have also empirically demonstrated its superior accuracy, stability and speed compared to other methods.

Several interesting directions remain for future research.  Many applications involve multivariate polytomous responses with more than two categories for each item. Extending SOLA to handle polytomous-response LCMs is an important future direction. It remains intriguing to investigate how to modify the algorithm accordingly and obtain similar theoretical guarantees. In addition, beyond estimation, practitioners often also want to conduct statistical inference on item parameters $\bTheta$. It is worth exploring  that whether we can build on SOLA to develop an estimator of item parameters $\bTheta$ together with confidence intervals.

% \paragraph{Supplementary Material.}
% The proofs of all theoretical results are included in the Supplementary Material.

\paragraph{Financial support.} This research is partially supported by the NSF Grant DMS-2210796.

\paragraph{Competing interests.} None.

% \clearpage
\appendix
\section{Appendix of the Proofs}

We present a general version of \Cref{prop:MLE-spectral-ratio-simple}. To this end, we introduce some additional notation. For $\forall j\in[J]$ and $k_1\ne k_2\in[K]$, define $\Delta_{j,k_1,k_2}^2:={\brac{\theta_{j,k_1}-\theta_{j,k_2}}^2}/\sqbrac{8(\sigma^2_{\theta_{j,k_1}}\vee \sigma^2_{\theta_{j,k_2}})}$.
 We have the following result, whose proof can be found in \Cref{pf-prop:MLE-spectral-ratio}.
\begin{proposition}\label{prop:MLE-spectral-ratio}
For any $k_1\ne k_2\in[K]$, define 
\begin{align*}
\tau_{k_1,k_2}:=\frac{\sum_{j\in\Omega_0\brac{k_1,k_2}}\Delta^2_{j,k_1,k_2}\cdot \tau \brac{\theta_{j,k_1}}}{\sum_{j\in\Omega_0\brac{k_1,k_2}}\Delta^2_{j,k_1,k_2}}\brac{1+\eta}, \quad  \tau(x)=\frac{1-2x}{2x(1-x)\log\brac{\brac{1-x}/x}},
\end{align*}
for some $\eta=o(1)$. 
Assume that (a)  $\max_{k_1\ne k_2\in[K]}\ab{\Omega^c_0(k_1,k_2)}=O(1)$ where $\Omega_0(k_1,k_2):=\ebrac{j\in[J]:\ab{\theta_{j,k_1}-\theta_{j,k_2}}=o(1)}$, then we have
\begin{align*}
\exp\brac{-\frac{I^\star}{2}} &\lesssim\exp\brac{-\frac{\Delta^2}{8\bar\sigma^2}\brac{1+\munderbar{\tau}}},
\end{align*}
where $\munderbar{\tau}:=\min_{k_1\ne k_2\in[K]}\tau_{k_1,k_2}$. Furthermore, if (b) $\min_{k\in[K]}\ab{\ebrac{j\in[J]:\ab{\theta_{j,k}-1/2}>c_0}}\gtrsim J$ for some  universal constant $c_0>0$, then   we have $\munderbar{\tau}>c_{\tau}$ for some universal constant $c_\tau>0$.
\end{proposition}

\section{Proofs of Main Results}
\subsection{Proof of Theorem \ref{thm:minimax-lower-bound} }\label{pf-thm:minimax-lower-bound} 
The proof  can be adapted from  Theorem S.4 in \cite{lyu2025degree} by fixing the individual degree parameters therein to be all $1$'s. We only sketch essential modifications here. Let $$\calP_\Theta:=\ebrac{\wt\bTheta:\wt\theta_{j,k}\in[c_\theta,C_\theta],j\in[J],k\in[K], \max_{j\in[J]}\max_{k\ne k^\prime} \ab{\wt\theta_{j,k}-\wt\theta_{j,k^\prime}}=o(1)}. $$ Following the construction of $\calV^*$ the proof  of Theorem S.4 in \cite{lyu2025degree}, it suffices to consider the following space  
\begin{align*}
    &\bar\calP_K\brac{\s,\bTheta}:=\ebrac{\bar\bR:\bar R_{i,j}=\theta_{j,s_i}, \s\in\calV^*,   \bTheta\in\calP_\Theta}.
\end{align*}
Then we can follow the proof  of Theorem S.4 in \cite{lyu2025degree}. Notice that in the proof of Lemma S.2, the argument therein still holds by allowing $p_{1,j}\asymp p_{2,j}\asymp 1$ and $\ab{p_{1,j}-p_{2,j}}=o\brac{p_{1,j}}$. We  can  then get the desired result by directly applying the modified Lemma S.2 without further lower bounding it.

\subsection{Proof of Theorem \ref{thm:refinemnet-err-split}}\label{pf-thm:refinemnet-err-split}
\paragraph{Initialization} We start with investigating the performance of  $\hat\theta^{(m)}_{j,k}$'s.  To this end, for any $\rho=o(1)$ we define the event
\begin{align}\label{eq:event-theta-error}
\calE_{\rho,m}:=\ebrac{\min_{\pi\in \frakG_K}\max_{j\in[J],k\in[K]} \ab{\hat\theta^{(m)}_{j,k}-\theta_{j,\pi\brac{k}}}\le \rho I^\star/J}.
\end{align}
The following lemma indicates that \eqref{eq:event-theta-error} holds with high probability, whose proof is deferred to \Cref{sec:proof-lem-init}.
\begin{lemma}\label{lem:init}
     There exists some absolute constant $c_0\in(0,1)$ such that $\PP\brac{\calE_{\rho,m}}\ge 1-\exp\brac{-\brac{1-c_0}\Delta^2/\brac{8\bar\sigma^2}}$ for $m\in[2]$ with $\rho=o(1/K)$.
\end{lemma}
% Without loss of generality, we assume that $\calS_1=\ebrac{1,\cdots,N/2}$ and $\calS_2=\ebrac{N/2+1,\cdots,N}$. 
\paragraph{One-step refinement} In the following, we will derive the clustering error for $i\in\calS_1$. Fix any $k\in[K]\backslash\ebrac{s_i^\star}$. Without loss of generality, we assume $\pi = \text{Id}$ (identity map with $\pi(k)=k$ for all $k$) in \eqref{eq:event-theta-error} as otherwise we can replace  $\theta_{j,k}$ with $\theta_{j,\pi(k)}$ and $\theta_{j,s_i^*}$ with $\theta_{j,\pi(s_i^*)}$ in the following derivation.  By definition we have
% We then investigate the performance of $\ebrac{\wt s_i}_{i\in[N]}$. For any $i\in[N]$ and $k\in[K]\backslash\ebrac{s^\star_i}$, we have that 
\begin{align}\label{eq:lov-event-mis-decomp}
    % &\II\brac{\hat s_i=k}\notag\\
\II\brac{\hat s_i=k}&\le \II\brac{\sum_{j\in[J]}R_{i,j}\log\frac{\hat\theta^{(2)}_{j,k}\brac{1-\hat\theta^{(2)}_{j,s^\star_i}}}{\hat\theta^{(2)}_{j,s^\star_i}\brac{1-\hat\theta_{j,k}}}\ge \sum_{j\in[J]}\log\frac{1-\hat\theta^{(2)}_{j,s^\star_i}}{1-\hat\theta^{(2)}_{j,k}}}\notag\\
&\le \II\brac{\sum_{j\in[J]}R_{i,j}\log\frac{\theta_{j,k}\brac{1-\theta_{j,s^\star_i}}}{\theta_{j,s^\star_i}\brac{1-\theta_{j,k}}}\ge \sum_{j\in[J]}\log\frac{1-\theta_{j,s^\star_i}}{1-\theta_{j,k}}-\delta I^\star}\notag\\
&+\II\left(\sum_{j\in[J]}R_{i,j}\brac{\log\frac{\hat\theta^{(2)}_{j,k}\brac{1-\hat\theta^{(2)}_{j,s^\star_i}}}{\hat\theta^{(2)}_{j,s^\star_i}\brac{1-\hat\theta^{(2)}_{j,k}}}-\log\frac{\theta_{j,k}\brac{1-\theta_{j,s^\star_i}}}{\theta_{j,s^\star_i}\brac{1-\theta_{j,k}}}}\right.\notag\\
&\ge \left.\sum_{j\in[J]}\brac{\log\frac{1-\hat\theta^{(2)}_{j,s^\star_i}}{1-\hat\theta^{(2)}_{j,k}}-\log\frac{1-\theta_{j,s^\star_i}}{1-\theta_{j,k}}}+\delta I^\star\right),
\end{align}
where $\delta=o(1)$ is some sequence to be determined later. We start with bounding the first term in \eqref{eq:lov-event-mis-decomp}. By Chernoff bound, we obtain that 
\begin{align}\label{eq:lov-ideal-delta-rate}
\PP&\brac{\sum_{j\in[J]}R_{i,j}\log\frac{\theta_{j,k}\brac{1-\theta_{j,s^\star_i}}}{\theta_{j,s^\star_i}\brac{1-\theta_{j,k}}}\ge \sum_{j\in[J]}\log\frac{1-\theta_{j,s^\star_i}}{1-\theta_{j,k}}-\delta I^\star}\notag\\
&\le \exp\brac{-\sum_{j\in[J]}\log\sqrt{\frac{1-\theta_{j,s^\star_i}}{1-\theta_{j,k}}}+\frac{\delta I^\star}{2}}\prod_{j=1}^J\brac{\sqrt{\theta_{j,s^\star_i}\theta_{j,k}}\sqrt{\frac{1-\theta_{j,s^\star_i}}{1-\theta_{j,k}}}+1-\theta_{j,s^\star_i}}\notag\\
&\le \exp\brac{\frac{\delta I^\star}{2}}\exp\brac{-\frac{I^\star}{2}}=\exp\brac{-\frac{I^\star}{2}\brac{1-\delta}}.
\end{align}
% We then bound the second term in \eqref{eq:lov-event-mis-decomp}. Notice that 
% \begin{align*}
%   &\ab{\sum_{j\in[J]}\brac{\log\frac{1-\hat\theta^{(2)}_{j,s^\star_i}}{1-\hat\theta^{(2)}_{j,k}}-\log\frac{1-\theta_{j,s^\star_i}}{1-\theta_{j,k}}}}\\
%   % \le \sum_{j\in[J]}\ab{\log\frac{1-\hat\theta^{(2)}_{j,s^\star_i}}{1-\theta_{j,s^\star_i}}}+\sum_{j\in[J]}\ab{\log\frac{1-\theta_{j,k}}{1-\hat\theta^{(2)}_{j,k}}}\\
%   &\le \sum_{j\in[J]}\ab{\log\brac{1+\frac{\theta_{j,s^\star_i}-\hat\theta^{(2)}_{j,s^\star_i}}{1-\theta_{j,s^\star_i}}}}+\sum_{j\in[J]}\ab{\log\brac{1+\frac{\theta_{j,k}-\hat\theta^{(2)}_{j,k}}{1-\theta_{j,k}}}}\\
%   &\lesssim \sum_{j\in[J]}\ab{\hat\theta^{(2)}_{j,s^\star_i}-\theta_{j,s^\star_i}}+\sum_{j\in[J]}\ab{\hat\theta^{(2)}_{j,k}-\theta_{j,k}}\le 2\rho I^\star,
% \end{align*}
% where the last inequality holds on $\calE_\rho$. 
Using this, we can bound the second term in \eqref{eq:lov-event-mis-decomp} as
\begin{align}\label{eq:lov-err-decomp}
&\II\brac{\sum_{j\in[J]}R_{i,j}\brac{\log\frac{\hat\theta^{(2)}_{j,k}\brac{1-\hat\theta^{(2)}_{j,s^\star_i}}}{\hat\theta^{(2)}_{j,s^\star_i}\brac{1-\hat\theta^{(2)}_{j,k}}}-\log\frac{\theta_{j,k}\brac{1-\theta_{j,s^\star_i}}}{\theta_{j,s^\star_i}\brac{1-\theta_{j,k}}}}\ge \sum_{j\in[J]}\brac{\log\frac{1-\hat\theta^{(2)}_{j,s^\star_i}}{1-\hat\theta^{(2)}_{j,k}}-\log\frac{1-\theta_{j,s^\star_i}}{1-\theta_{j,k}}}+\delta I^\star}\notag\\
&\le \II\brac{\sum_{j\in[J]}R_{i,j}\log\frac{\hat\theta^{(2)}_{j,k}}{\theta_{j,k}}>\frac{\delta}{4}I^\star}+\II\brac{\sum_{j\in[J]}R_{i,j}\log\frac{\theta_{j,s^\star_i}}{\hat \theta_{j,s^\star_i}}>\frac{\delta}{4}I^\star}\notag\\
&+\II\brac{\sum_{j\in[J]}\brac{R_{i,j}-1}\log\frac{1-\hat\theta^{(2)}_{j,s^\star_i}}{1-\theta_{j,s^\star_i}}>\frac{\delta}{4}I^\star}+\II\brac{\sum_{j\in[J]}\brac{R_{i,j}-1}\log\frac{1-\theta_{j,k}}{1-\hat\theta^{(2)}_{j,k}}>\frac{\delta}{4}I^\star}.
\end{align}
For the first term in \eqref{eq:lov-err-decomp}, we proceed by noticing that 
\begin{align}\label{eq:lov-err-decomp-2}
&\II\brac{\sum_{j\in[J]}R_{i,j}\log\frac{\hat\theta^{(2)}_{j,k}}{\theta_{j,k}}>\frac{\delta}{4}I^\star}\notag\\
&\le \II\brac{\sum_{j\in[J]}R_{i,j}\log\frac{\hat\theta^{(2)}_{j,k}}{\theta_{j,k}}\II\brac{\calE_{\rho,2}}>\frac{\delta}{8}I^\star}+\II\brac{\sum_{j\in[J]}R_{i,j}\log\frac{\hat\theta^{(2)}_{j,k}}{\theta_{j,k}}\II\brac{\calE^c_{\rho,2}}>\frac{\delta}{8}I^\star}.
\end{align}
Let $\delta=\rho^{\epsilon}$ for $\epsilon\in(0,1)$ and $\lambda=c\rho^{-\epsilon}$ for some constant $c>0$ depending only on $c_\theta$ and $C_\theta$, then by Chernoff bound on the first term of \eqref{eq:lov-err-decomp-2} we obtain that
\begin{align*}
&\PP\brac{\sum_{j\in[J]}R_{i,j}\log\frac{\hat\theta^{(2)}_{j,k}}{\theta_{j,k}}\II\brac{\calE_{\rho,2}}>\frac{\delta}{8}I^\star}\overset{(a)}{\le} \PP\brac{\sum_{j\in[J]}R_{i,j}\frac{\rho I^\star}{J}>\frac{c_\theta\delta}{8}I^\star}\\
   &\le \exp\brac{-\lambda\frac{c_\theta\delta}{8}I^\star}\prod_{j=1}^J\EE \exp\brac{\lambda R_{i,j}\frac{\rho I^\star}{J}}
   \overset{(b)}{\le} \exp\brac{-\lambda\frac{c_\theta\delta}{8}I^\star}\prod_{j=1}^J\brac{1+2\lambda\theta_{i,j}\frac{\rho I^\star}{J}}\\
    &\le \exp\brac{-\lambda\frac{c_\theta\delta}{8}I^\star}\exp\brac{\rho \cdot 2C_\theta I^\star}\overset{(c)}{\le} \exp\brac{- \frac{ I^\star}{2}},
\end{align*}
where (a) holds due to \eqref{eq:event-theta-error} and $\log\brac{1+x}\le x$ for $x>-1$, (b) holds due to $e^{x}-1\le 2x$ for $0<x<1$ and the fact that ${\lambda \rho I^\star}/{J}\lesssim{\rho^{1-\epsilon}\Delta^2}/{J}=o(1)$, and (c) holds due to the choice of $\lambda=c\rho^{-\varepsilon}$. For the second term of \eqref{eq:lov-err-decomp-2}, by the independence between $\hat\theta^{(2)}_{j,k}$ and $\ebrac{R_{i,j},j\in[J]}$ we get  
\begin{align*}
    \PP&\brac{\sum_{j\in[J]}R_{i,j}\log\frac{\hat\theta^{(2)}_{j,k}}{\theta_{j,k}}\II\brac{\calE^c_{\rho,2}}>\frac{\delta}{8}I^\star}\le \exp\brac{-\lambda\frac{\delta}{8}I^\star}\prod_{j=1}^J\EE \exp\brac{\lambda R_{i,j}\log\frac{\hat\theta^{(2)}_{j,k}}{\theta_{j,k}}\II\brac{\calE^c_{\rho,2}}}\\
    &\le \exp\brac{-\lambda\frac{\delta}{8}I^\star}\prod_{j=1}^J\sqbrac{1+\theta_{j,k}\brac{\EE \exp\brac{\lambda \log\frac{\hat\theta^{(2)}_{j,k}}{\theta_{j,k}}\II\brac{\calE^c_{\rho,2}}}-1}}.
\end{align*}
To bound the above term, notice that
\begin{align*}
    \EE \brac{\exp\brac{\lambda \log\frac{\hat\theta^{(2)}_{j,k}}{\theta_{j,k}}\II\brac{\calE^c_{\rho,2}}}-1}\le \brac{\exp\brac{\lambda \log \frac{1}{c_\theta}}-1}\PP\brac{\calE^c_{\rho,2}}\le \exp\brac{\lambda \log \frac{1}{c_\theta}-\frac{\Delta^2}{16\bar\sigma^2}},
\end{align*}
Since we can always choose $\rho\rightarrow 0$ sufficiently slow such that
\begin{align*}
    \lambda\log\frac{1}{c_\theta}\lesssim \frac{1}{\rho^{\varepsilon}}\lesssim\frac{\Delta^2}{\bar\sigma^2}\asymp I^\star,
\end{align*}
we can conclude that 
\begin{align*}
    \PP&\brac{\sum_{j\in[J]}R_{i,j}\log\frac{\hat\theta^{(2)}_{j,k}}{\theta_{j,k}}\II\brac{\calE^c_{\rho,2}}>\frac{\delta}{8}I^\star}\le \exp\brac{-\lambda\frac{\delta}{8}I^\star}\prod_{j=1}^J\sqbrac{1+\theta_{j,k}\exp\brac{-c{I^\star}}}\\
    &\le \exp\brac{-\lambda\frac{\delta}{8}I^\star}\brac{1+\frac{C_\theta}{\exp\brac{c{I^\star}}}}^J\le\exp\brac{-\frac{I^\star}{2}}.
\end{align*}
where  the last inequality due to  the fact that $ \brac{1+C_\theta{\exp\brac{-c{I^\star}}}}^J=O(1)$ provided that $\exp\brac{c{I^\star}}\ge  J$. We thereby arrive at
\begin{align}\label{eq:lov-err-decomp-2-bound}
\PP\brac{\sum_{j\in[J]}R_{i,j}\log\frac{\hat\theta^{(2)}_{j,k}}{\theta_{j,k}}>\frac{\delta}{4}I^\star}\le 2\exp\brac{-\frac{I^\star}{2}}.
\end{align}
The second term in \eqref{eq:lov-err-decomp} can bounded in the same fashion. For the third term in \eqref{eq:lov-err-decomp}, we have 
\begin{align*}
    &\II\brac{\sum_{j\in[J]}\brac{R_{i,j}-1}\log\frac{1-\hat\theta^{(2)}_{j,s^\star_i}}{1-\theta_{j,s^\star_i}}>\frac{\delta}{4}I^\star}\notag\\
    &\le\II\brac{\sum_{j\in[J]}\brac{R_{i,j}-1}\log\frac{1-\hat\theta^{(2)}_{j,s^\star_i}}{1-\theta_{j,s^\star_i}}\II\brac{\calE_{\rho,2}}>\frac{\delta}{8}I^\star}+\II\brac{\sum_{j\in[J]}\brac{R_{i,j}-1}\log\frac{1-\hat\theta^{(2)}_{j,s^\star_i}}{1-\theta_{j,s^\star_i}}\II\brac{\calE^c_{\rho,2}}>\frac{\delta}{8}I^\star}.    
\end{align*}
Similarly, we can bound the expectation of the first term in the above equation as
\begin{align*}
    &\PP\brac{\sum_{j\in[J]}\brac{R_{i,j}-1}\log\frac{1-\hat\theta^{(2)}_{j,s^\star_i}}{1-\theta_{j,s^\star_i}}\II\brac{\calE_{\rho,2}}>\frac{\delta}{8}I^\star}\\
    &\le \exp\brac{-\lambda\frac{\brac{1-C_{\theta}}\delta}{8}I^\star}\prod_{j=1}^J\EE \exp\brac{\lambda \brac{R_{i,j}-1}\frac{\rho I^*}{J}}\\
    % &= \exp\brac{-\lambda\frac{\brac{1-C_{\theta}}\delta}{8}I^\star}\prod_{j=1}^J\brac{\theta_{j,s^\star_i}+\brac{1-\theta_{j,s^\star_i}}\exp\brac{-\frac{\rho I^*}{J}}}\\
    &\overset{(a)}{\le} \exp\brac{-\lambda\frac{\brac{1-C_{\theta}}\delta}{8}I^\star}\prod_{j=1}^J\brac{1+\frac{C_{\theta}\rho I^*}{2J}}\\
    &\le \exp\brac{-\lambda\frac{\brac{1-C_{\theta}}\delta}{8}I^\star}\exp\brac{\rho \cdot \frac{C_\theta I^*}{2}}\le \exp\brac{-\frac{I^\star}{2}},
\end{align*}
where (a) holds due to $e^{-x}\le 1-x/2$ for $x\in[0,1]$. Similarly we have that
\begin{align*}
    &\PP\brac{\sum_{j\in[J]}\brac{R_{i,j}-1}\log\frac{1-\hat\theta^{(2)}_{j,s^\star_i}}{1-\theta_{j,s^\star_i}}\II\brac{\calE^c_{\rho,2}}>\frac{\delta}{8}I^\star}\\
    &\le \exp\brac{-\lambda\frac{\delta}{8}I^\star}\prod_{j=1}^J\sqbrac{\theta_{j,s_i^*}+\brac{1-\theta_{j,s_i^*}}\EE \exp\brac{\lambda \log\frac{1-\hat\theta^{(2)}_{j,s_i^*}}{1-\theta_{j,s_i^*}}\II\brac{\calE^c_{\rho,2}}}}\\
    &\le \exp\brac{-\lambda\frac{\delta}{8}I^\star}\prod_{j=1}^J\sqbrac{\theta_{j,s_i^*}+\brac{1-\theta_{j,s_i^*}} \exp\brac{\lambda \log\frac{1}{1-C_{\theta}}}\PP\brac{\calE^c_{\rho,2}}}\\
    &\le \exp\brac{-\lambda\frac{\delta}{8}I^\star}\prod_{j=1}^J\sqbrac{\theta_{j,s_i^*}+\brac{1-\theta_{j,s_i^*}} \exp\brac{-cI^*}}\\
    &\le \exp\brac{-\lambda\frac{\delta}{8}I^\star}C_{\theta}^J\brac{1+\frac{1}{c_{\theta}\exp\brac{cI^*}}}^J\le \exp\brac{-\frac{I^\star}{2}}.
\end{align*}
provided that $\exp\brac{c{I^\star}}\ge  J$.  We thus get that
\begin{align}\label{eq:lov-err-decomp-3-bound}
    \PP\brac{\sum_{j\in[J]}\brac{R_{i,j}-1}\log\frac{1-\hat\theta^{(2)}_{j,s^\star_i}}{1-\theta_{j,s^\star_i}}>\frac{\delta}{4}I^\star}\le 2\exp\brac{-\frac{I^\star}{2}}.
\end{align}
The last term in \eqref{eq:lov-err-decomp} can be bounded in the same fashion.  By \eqref{eq:lov-event-mis-decomp}, \eqref{eq:lov-ideal-delta-rate} \eqref{eq:lov-err-decomp}, \eqref{eq:lov-err-decomp-2-bound} and \eqref{eq:lov-err-decomp-3-bound}, we thereby arrive at
\begin{align*}
\PP\brac{\hat s_i=k}&\le \exp\brac{-\frac{I^\star}{2}\brac{1-\rho^\epsilon}}+8\exp\brac{-\frac{I^\star}{2}}\le \exp\brac{-\frac{I^\star}{2}\brac{1-o(1)}}.
\end{align*}
We thus conclude that 
\begin{align*}
   \frac{1}{\ab{\calS_1}} \EE\sum_{i\in\calS_1}\II\brac{\hat s_i\ne  s^\star_i}\le \exp\brac{-\frac{I^\star}{2}\brac{1-o(1)}}.
\end{align*}
The proof for the case $i\in\calS_2$ can be conducted in the same fashion. We can conclude that
\begin{align*}
    \frac{1}{\ab{\calS_2}} \EE\sum_{i\in\calS_2}\II\brac{\hat s_i\ne  \pi_0\brac{s^\star_i}}\le \exp\brac{-\frac{I^\star}{2}\brac{1-o(1)}}.
 \end{align*}
for some $\pi_0\in\frakG_K$. 
\paragraph{Label Alignment}
We then proceed with the proof for label alignment.  Recall that without loss of generality we have assumed that $ \ell\brac{\wt \s^{(1)},\s^{\star}_{\calS_1}}= \ell_0\brac{\wt \s^{(1)},\s^{\star}_{\calS_1}}$ and $ \ell\brac{\wt \s^{(2)},\s^{\star}_{\calS_2}}= \ell_0\brac{\wt \s^{(2)},\pi_0(\s^{\star}_{\calS_2})}$ for some $\pi_0\in\frakG_K$. Notice that
\begin{align*}
    \ell_0\brac{\hat \s_{\calS_2},\s^{\star}_{\calS_2}}&=\ell_0\brac{\hat\pi(\hat \s^{(2)}),\s^{\star}_{\calS_2}}=\ell_0\brac{\hat \s^{(2)},\hat\pi^{-1}(\s^{\star}_{\calS_2})}
\end{align*}
It suffices to show that $\hat\pi=\pi_0^{-1}$, then we have
\begin{align*}
    \EE\ell_0\brac{\hat \s^{(2)},\hat\pi^{-1}(\s^{\star}_{\calS_2})}=\EE\ell_0\brac{\hat \s^{(2)},\pi_0(\s^{\star}_{\calS_2})}\le \exp\brac{-\frac{I^\star}{2}\brac{1-o(1)}},
\end{align*}
as desired.  By definition, we have
\begin{align*}
\fro{\hat\bTheta^{(1)}-\hat\bTheta^{(2)}\bG_{\pi_0^{-1}}}^2&=\sum_{j\in[J]}\sum_{k\in[K]}\ab{\hat\theta^{(1)}_{j,k}-\hat\theta_{j,\pi_0^{-1}(k)}^{(2)}}^2\\
&\le \sum_{j\in[J]}\sum_{k\in[K]}\ab{\hat\theta^{(1)}_{j,k}-\theta_{j,k}}^2+ \sum_{j\in[J]}\sum_{k\in[K]}\ab{\hat\theta^{(2)}_{j,\pi_0^{-1}(k)}-\theta_{j,k}}^2\le 2\rho KI^\star
\end{align*}
On the other hand, for any $\pi\in\frakG_K$ and $\pi\ne \pi_0^{-1}$ we have
\begin{align*}
    \fro{\hat\bTheta^{(1)}-\hat\bTheta^{(2)}\bG_{\pi}}^2&=\sum_{j\in[J]}\sum_{k\in[K]}\ab{\hat\theta^{(1)}_{j,k}-\hat\theta_{j,\pi(k)}^{(2)}}^2\\
    & \ge \sum_{j\in[J]}\sum_{k\in[K]}\ab{\hat\theta_{j,\pi(k)}^{(2)}-\theta_{j,k}}^2-\rho KI^\star
    \end{align*}
% Combining the SNR condition and the proof of \Cref{lem:init}  gives that on $\calE_{\rho,2}$,
% \begin{align*}
%     \ell\brac{\wt \s^{(2)},\s^{\star}_{\calS_2}}= \ell_0\brac{\wt \s^{(2)},\pi_0(\s^{\star}_{\calS_2})}\le \exp\brac{-\brac{1-c_0}\frac{\Delta^2}{8}}=o\brac{\frac{1}{K}}.
% \end{align*}
% Since $\pi\ne \pi_0$, we thus have $\ell_0\brac{\wt \s^{(2)},\pi(\s^{\star}_{\calS_2})}>\alpha/K$. Recall the definition of $\theta_{j,k}\brac{\s}$, $\bar\theta_{j,k}\brac{\s}$ and $\frakE_{j,k}\brac{\s}$ in \Cref{sec:proof-theta-err}. 
It suffices to show that $ \sum_{j\in[J]}\sum_{k\in[K]}\ab{\hat\theta_{j,\pi(k)}^{(2)}-\theta_{j,k}}^2>4\rho K I^\star$. Observe that
\begin{align*}
    \ab{\hat\theta_{j,\pi(k)}^{(2)}-\theta_{j,k}}&=\ab{\hat\theta_{j,\pi(k)}^{(2)}-\theta_{j,\pi\circ \pi_0(k)}+\theta_{j,\pi\circ \pi_0(k)}-\theta_{j,k}}\\
    &\ge \ab{\theta_{j,\pi\circ \pi_0(k)}-\theta_{j,k}}-\ab{\hat\theta_{j,\pi(k)}^{(2)}-\theta_{j,\pi\circ \pi_0(k)}}.
\end{align*} 
Notice that $\ab{\hat\theta_{j,\pi(k)}^{(2)}-\theta_{j,\pi\circ \pi_0(k)}}=\ab{\hat\theta_{j,k}^{(2)}-\theta_{j,\pi_0(k)}}\le \rho I^\star/J$, we have
\begin{align*}
    &\sum_{j=1}^J\sum_{k=1}^K\brac{\hat\theta_{j,\pi(k)}^{(2)}-\theta_{j,k}}^2\\
    &\ge \sum_{j=1}^J\sum_{k=1}^K\brac{\theta_{j,\pi\circ \pi_0(k)}-\theta_{j,k}}^2-2\sum_{j=1}^J\sum_{k=1}^K\ab{\theta_{j,\pi\circ \pi_0(k)}-\theta_{j,k}}\ab{\hat\theta_{j,\pi(k)}^{(2)}-\theta_{j,\pi\circ \pi_0(k)}}\\
    &\ge \sum_{j=1}^J\sum_{k=1}^K\brac{\theta_{j,\pi\circ \pi_0(k)}-\theta_{j,k}}^2-\frac{2\rho I^\star}{J}\sum_{j=1}^J\sum_{k=1}^K\ab{\theta_{j,\pi\circ \pi_0(k)}-\theta_{j,k}}\\
    &\ge \brac{\sum_{k=1}^K\II\brac{\pi\circ\pi_0(k)\ne k}} \brac{\Delta^2-4C_\theta\rho I^\star}\ge \Delta^2
\end{align*}
where the last inequality holds due to the fact that $\pi\ne \pi_0^{-1}$, $\Delta^2\asymp I^\star\rightarrow \infty$ and $\rho=o(1)$. The proof is thus completed by using $\rho=o\brac{1/K}$.

\subsection{Proof of Proposition \ref{prop:lam-delta}}
Notice that $\lambda_K\ge \lambda_{K}\brac{\bZ}\lambda_{K}\brac{\bTheta}$. To get a lower bound for $\lambda_K$, it suffices to obtain lower bounds for $\lambda_{K}\brac{\bZ}$ and $\lambda_{K}\brac{\bTheta}$. First note that 
\begin{align*}
\bZ^\top\bZ=\text{diag}\brac{\ab{\ebrac{i:s_i^\star=1}},\cdots,\ab{\ebrac{i:s_i^\star=K}}}.
\end{align*}
We thus obtain  $\lambda_{K}\brac{\bZ}\ge \sqrt{\alpha N/K}$. On the other hand, by Proposition 2 in \cite{lyu2025degree} we obtain   that for any $\delta\in(0,1)$,
\begin{align*}
    \PP\brac{\lambda_K\brac{\bTheta^\top\bTheta}\ge \brac{1-\delta} BJK}\le K\brac{e^{-\delta^2B^2J/16}+e^{-3\delta BJ/8}}.
\end{align*}
Denote the above event as $\calE_{\theta}$, under which we arrive at $\lambda_K\brac{\bTheta}\ge \brac{1-\delta/2}\sqrt{BJK}$. 

To obtain an lower bound for $\Delta$, it suffices to use Lemma S.1 in \cite{lyu2025degree} to get $\Delta\ge \sqrt{2}\lambda_K\brac{\bTheta}$. We get the desired result by setting $\delta=1/2$.

\subsection{Proof of Corollary \ref{thm:cem-refinemnet-err-split}}\label{sec:pf-thm:cem-refinemnet-err-split}
The proof follows the same argument as that of \Cref{thm:refinemnet-err-split}. We only sketch the modifications here. Without loss of generality, we only consider sample data points in $\calS_1$. First note that by Proposition 3.1 in \cite{zhang2022leave}, we have $\ell\brac{\wt\s^{(2)},\s_{\calS_2}^\star}=o(1)$ with probability at least $1-O\brac{e^{-N\vee J}}$. Hence we get that 
\begin{align*}
    \log\brac{\frac{\hat  p_k}{\hat \pi_{s_i^\star}}}=\log\brac{\frac{\hat  p_k}{ p_k}}+\log\brac{\frac{ \pi_{s_i^\star}}{ \hat \pi_{s_i^\star}}}+\log\brac{\frac{ p_k}{\pi_{s_i^\star}}}\lesssim \log\brac{\frac{1}{\alpha}}=o\brac{I},
\end{align*}
with probability at least $1-O\brac{e^{-N\vee J}}$, where we've used \Cref{assump:balanced}. This indicates that the impact of this additional term is ignorable, i.e.,
\begin{align*}
    \PP\brac{ \log\brac{\frac{\hat  p_k}{\hat \pi_{s_i^\star}}}>{c\delta}I^\star}\le O\brac{e^{-N\vee J}},
\end{align*}
for some universal constant $c>0$. The remaining proof follows from that of \Cref{thm:refinemnet-err-split} line by line.

\subsection{Proof of Proposition \ref{prop:MLE-spectral-ratio}}\label{pf-prop:MLE-spectral-ratio}
Specifically, we fix $j\in\Omega_0(k_1,k_2)$ and $k_1\ne k_2\in[K]$ in the following analysis. Note that 
\begin{align*}
&-2\log\brac{\sqrt{\theta_{j,k_1}\theta_{j,k_2}}+\sqrt{\brac{1-\theta_{j,k_1}}\brac{1-\theta_{j,k_2}}}}\\
&=-\log\brac{1-\brac{\sqrt{\theta_{j,k_1}}-\sqrt{\theta_{j,k_2}}}^2-2\sqrt{\theta_{j,k_1}\theta_{j,k_2}}\brac{1-\sqrt{\theta_{j,k_1}\theta_{j,k_2}}-\sqrt{\brac{1-\theta_{j,k_1}}\brac{1-\theta_{j,k_2}}}}}\\
&\ge \brac{\sqrt{\theta_{j,k_1}}-\sqrt{\theta_{j,k_2}}}^2+2\sqrt{\theta_{j,k_1}\theta_{j,k_2}}\brac{1-\sqrt{\theta_{j,k_1}\theta_{j,k_2}}-\sqrt{\brac{1-\theta_{j,k_1}}\brac{1-\theta_{j,k_2}}}}
\end{align*}
When $\theta_{j,k_1}+\theta_{j,k_2}\le 1$, we have
\begin{align*}
    1-\sqrt{\theta_{j,k_1}\theta_{j,k_2}}-\sqrt{\brac{1-\theta_{j,k_1}}\brac{1-\theta_{j,k_2}}}&=\frac{\brac{1-\sqrt{\theta_{j,k_1}\theta_{j,k_2}}}^2-\brac{1-\theta_{j,k_1}}\brac{1-\theta_{j,k_2}}}{1-\sqrt{\theta_{j,k_1}\theta_{j,k_2}}+\sqrt{\brac{1-\theta_{j,k_1}}\brac{1-\theta_{j,k_2}}}}\\
    &=\frac{\brac{\sqrt{\theta_{j,k_1}}-\sqrt{\theta_{j,k_2}}}^2}{1-\sqrt{\theta_{j,k_1}\theta_{j,k_2}}+\sqrt{\theta_{j,k_1}\theta_{j,k_2}+1-\theta_{j,k_1}-\theta_{j,k_2}}}\\
    &=C_{\theta_{j,k_1},\theta_{j,k_2}}\brac{\sqrt{\theta_{j,k_1}}-\sqrt{\theta_{j,k_2}}}^2
\end{align*}
When $\theta_{j,k_2}+\theta_{j,k_1}>1$, we have that 
\begin{align*}
    1-\sqrt{\theta_{j,k_1}\theta_{j,k_2}}-\sqrt{\brac{1-\theta_{j,k_1}}\brac{1-\theta_{j,k_2}}}&=\frac{\brac{1-\sqrt{\brac{1-\theta_{j,k_1}}\brac{1-\theta_{j,k_2}}}}^2-\theta_{j,k_1}\theta_{j,k_2}}{1-\sqrt{\brac{1-\theta_{j,k_1}}\brac{1-\theta_{j,k_2}}}+\sqrt{\theta_{j,k_1}\theta_{j,k_2}}}\\
    &=\frac{\brac{\sqrt{1-\theta_{j,k_1}}-\sqrt{1-\theta_{j,k_2}}}^2}{1+\sqrt{\theta_{j,k_1}\theta_{j,k_2}}-\sqrt{\theta_{j,k_1}\theta_{j,k_2}-\brac{\theta_{j,k_1}+\theta_{j,k_2}-1}}}\\
    &=C_{\theta_{j,k_1},\theta_{j,k_2}}\brac{\sqrt{1-\theta_{j,k_1}}-\sqrt{1-\theta_{j,k_2}}}^2.
\end{align*}
Here for any $\theta_1,\theta_2\in[0,1]$ we define
\begin{align*}
    C_{\theta_1,\theta_2}:=\begin{cases}
        \brac{1-\sqrt{\theta_{1}\theta_{2}}+\sqrt{\theta_{1}\theta_{2}+1-\theta_{1}-\theta_{2}}}^{-1}, & \theta_1+\theta_2\le 1\\
        \brac{1+\sqrt{\theta_{1}\theta_{2}}-\sqrt{\theta_{1}\theta_{2}-\brac{\theta_{1}+\theta_{2}-1}}}^{-1}, & \theta_1+\theta_2> 1
    \end{cases}.
\end{align*}
In addition, let $\bar C_{\theta_1,\theta_2}:=2C_{\theta_1,\theta_2}\sqrt{\theta_1\theta_2}$. Hence we conclude that 
\begin{align*}
-2\log\brac{\sqrt{\theta_{j,k_1}\theta_{j,k_2}}+\sqrt{\brac{1-\theta_{j,k_1}}\brac{1-\theta_{j,k_2}}}}\ge\brac{1+\bar C_{\theta_{j,k_1},\theta_{j,k_2}}}\textsf{D}_2\brac{\theta_{j,k_1},\theta_{j,k_2}}.
\end{align*}
where for any $\theta_1,\theta_2\in[0,1]$, we define
\begin{align*}
\textsf{D}_2\brac{\theta_1,\theta_2}:=\begin{cases}
        \brac{\sqrt{\theta_1}-\sqrt{\theta_2}}^2, & \theta_1+\theta_2\le 1\\
         \brac{\sqrt{1-\theta_1}-\sqrt{1-\theta_2}}^2, &
         \theta_1+\theta_2> 1
    \end{cases}.
\end{align*}
By the definition of $\Omega_0(k_1,k_2)$, we have either (i) $\theta_{j,k_1}\vee\theta_{j,k_2}< 1/2$ and $\theta_{j,k_1}+\theta_{j,k_2}\le  1$, or (ii) $\theta_{j,k_1}\wedge\theta_{j,k_2}> 1/2$ and $\theta_{j,k_1}+\theta_{j,k_2}> 1$.  

We start with the case (i).  Without loss of generality we assume $\frac{1}{2}>\theta_{j,k_1}>\theta_{j,k_2}$ and denote $\theta:=\theta_2$, $\delta:=\theta_1-\theta_2=o(1)$.
By definition we have
\begin{align*}
    -\log\brac{\sqrt{\theta_{j,k_1}\theta_{j,k_2}}+\sqrt{\brac{1-\theta_{j,k_1}}\brac{1-\theta_{j,k_2}}}}&=\frac{\brac{1+\bar C_{\theta+\delta,\theta}}}{2}\brac{\sqrt{\theta+\delta}-\sqrt{\theta}}^2\brac{1+o(1)}\\
    &=\delta^2\frac{\brac{1+\bar C_{\theta+\delta,\theta}}}{8\theta}\brac{1+o(1)}\\
    &=\frac{\delta^2}{8\theta\brac{1-\theta}}\brac{1+o(1)},
\end{align*}
On the other hand, we have
\begin{align*}
    \frac{\brac{\theta_{j,k_1}-\theta_{j,k_2}}^2}{8\brac{\sigma^2_{\theta_{j,k_1}}\vee \sigma^2_{\theta_{j,k_1}}}}=\delta^2\frac{\log\brac{\brac{1-\theta}/\theta}}{4\brac{1-2\theta}}
\end{align*}
This implies that 
\begin{align*}
   \frac{-\log\brac{\sqrt{\theta_{j,k_1}\theta_{j,k_2}}+\sqrt{\brac{1-\theta_{j,k_1}}\brac{1-\theta_{j,k_2}}}}}{\brac{\theta_{j,k_1}-\theta_{j,k_2}}^2/\brac{8\brac{\sigma^2_{\theta_{j,k_1}}\vee \sigma^2_{\theta_{j,k_1}}}}}=\frac{1-2\theta}{2\theta\brac{1-\theta}\log\brac{\brac{1-\theta}/\theta}}(1+o(1))
\end{align*}
Define the function
\begin{align*}
    \tau(x)=\frac{1-2x}{2x(1-x)\log\brac{\brac{1-x}/x}},\quad g(x)=\brac{-2x^2+2x-1}\log\brac{\frac{1}{x}-1}-2x+1,
\end{align*}
for $0<x<{1}/{2}$. Some algebra gives that
\begin{align*}
    \tau^{\prime}\brac{x}=\frac{g(x)}{2x^2\brac{1-x}^2\log^2\brac{\frac{1}{x}-1}}
\end{align*}
Note that 
\begin{align*}
    g^\prime(x)=\frac{\brac{1-2x}^2}{x(1-x)}+2\brac{1-2x}\log\brac{\frac{1}{x}-1}>0,
\end{align*}
for $0<x<{1}/{2}$, which implies that $g(x)< \lim_{x\rightarrow \frac{1}{2}} g(x)=0$. Hence  we have $\tau^\prime(x)< 0$ and  $\tau(x)> \lim_{x\rightarrow \frac{1}{2}} \tau(x)=1$.
Thus we have
\begin{align*}
   \frac{-\log\brac{\sqrt{\theta_{j,k_1}\theta_{j,k_2}}+\sqrt{\brac{1-\theta_{j,k_1}}\brac{1-\theta_{j,k_2}}}}}{\brac{\theta_{j,k_1}-\theta_{j,k_2}}^2/\brac{8\brac{\sigma^2_{\theta_{j,k_1}}\vee \sigma^2_{\theta_{j,k_1}}}}}=\tau(\theta)(1+o(1))>1,
\end{align*}
where $\tau\brac{\theta}$ is monotone decreasing as $\theta$ increases to $1/2$ and $\tau\brac{\theta}\rightarrow \infty$ as $\theta\rightarrow 0$.

The analysis for case (ii) is essentially the same and hence omitted. Combining (i) and (ii), we can arrive at
\begin{align*}
I_{k_1,k_2}=\sum_{j\in\Omega_0\brac{k_1,k_2}}I\brac{\theta_{j,k_1},\theta_{j,k_2}}+\sum_{j\in\Omega_0^c\brac{k_1,k_2}}I\brac{\theta_{j,k_1},\theta_{j,k_2}}
\end{align*}
and
\begin{align*}
\exp\brac{-\frac{I_{k_1,k_2}}{2}}&=\exp\brac{-\frac{1}{2}\sum_{j\in\Omega_0\brac{k_1,k_2}}I\brac{\theta_{j,k_1},\theta_{j,k_2}}}\exp\brac{-\frac{1}{2}\sum_{j\in\Omega_0^c\brac{k_1,k_2}}I\brac{\theta_{j,k_1},\theta_{j,k_2}}}\\
&\gtrsim\exp\brac{-\sum_{j\in\Omega_0\brac{k_1,k_2}}\Delta^2_{j,k_1,k_2}\cdot \frac{I\brac{\theta_{j,k_1},\theta_{j,k_2}}/2}{\Delta^2_{j,k_1,k_2}}}\\
&\gtrsim\exp\brac{-\frac{\sum_{j\in[J]}\brac{\theta_{j,k_1}-\theta_{j,k_2}}^2}{8\bar\sigma^2}\brac{1+\tau_{k_1,k_2}}}
\end{align*}
where the inequality holds due to the assumption $\ab{\Omega^c_0}=O(1)$, and
\begin{align*}
\tau_{k_1,k_2}:=\frac{\sum_{j\in\Omega_0\brac{k_1,k_2}}\Delta^2_{j,k_1,k_2}\cdot \tau \brac{\theta_{j,k_1}}}{\sum_{j\in\Omega_0\brac{k_1,k_2}}\Delta^2_{j,k_1,k_2}}\brac{1+o(1)}
\end{align*}

\subsection{Proof of Lemma \ref{lem:K-est}}\label{sec:proof-K-est}
In light of 
Corollary 3.12, Remark 3.13 in \cite{bandeira2016sharp} and Theorem 6.10 in \cite{boucheron2013concentration}, for $t\ge 0$ we have 
\begin{align*}
    \PP\brac{\op{\bE}\ge \frac{(1+\varepsilon)\sqrt{2}}{2}\brac{\sqrt{N}+\sqrt{J}}+C_{\varepsilon}\sqrt{\log\brac{N\wedge J}}+t}\le (N\wedge J)e^{-t^2/2}.
\end{align*}
By choosing $t=2\sqrt{\log \brac{N\wedge J}}$ and $\varepsilon=\sqrt{2}-1$, for $N,J$ large enough we obtain that 
\begin{align*}
    \PP\brac{\op{\bE}\ge 2\brac{\sqrt{N}+\sqrt{J}}}\le (N\wedge J)^{-1}.
\end{align*}
This implies that $ \lambda_{K+1}\brac{\bR}\le \op{\bE}\le  2\brac{\sqrt{J}+\sqrt{N}}$ with  probability $1-o(1)$. On the other hand, by assumption in \Cref{prop:spec-clustering} on $\sigma_{K}\brac{\EE \bR}$, we obtain that 
\begin{align*}
    \lambda_{K}\brac{\bR}\ge  \lambda_{K}\brac{\EE \bR}-\op{\bE}=\omega\brac{\sqrt{N}+\sqrt{J}},
\end{align*}
with  probability $1-o(1)$. This completes the proof of Lemma \ref{lem:K-est}.

\subsection{Special case when there exists $\theta_{j,k}$ close to $1/2$.}
For $\theta<1/2$ we have
\begin{align*}
    \sigma^2_\theta=\frac{1-2\theta}{2\log\brac{\brac{1-\theta}/\theta}}=\brac{\frac{1}{2}-\theta}\frac{1}{\log\brac{1+\frac{1}{\theta}-2}}\ge \frac{\theta}{2}.
\end{align*}
For $\theta>1/2$ we have
\begin{align*}
    \sigma^2_\theta=\frac{2\theta-1}{2\log\brac{\theta/\brac{1-\theta}}}=\brac{\theta-\frac{1}{2}}\frac{1}{\log\brac{1+\frac{2\theta-1}{1-\theta}}}\ge \frac{1-\theta}{2}.
\end{align*}
Denote $\bar\theta:=\argmin_{\theta\in\ebrac{\theta_{j,k},j\in[J], k\in[K]}}\ab{\frac{1}{2}-\theta}$ and $\bar\delta:=\ab{\frac{1}{2}-\bar\theta}$, then we simply have
\begin{align*}
\bar\sigma^2=\sigma^2_{\bar\theta}\ge \frac{\bar\theta\wedge\brac{1-\bar\theta}}{2}=\frac{1}{4}\brac{1-2\bar\delta}.
\end{align*}
We thereby have
\begin{align*}
    \frac{1}{8\bar\sigma^2}\brac{\theta_{j,k_1}-\theta_{j,k_2}}^2\le \frac{1}{2\brac{1-2\bar\delta}}\brac{\theta_{j,k_1}-\theta_{j,k_2}}^2
    % =2\brac{\brac{\frac{\theta_{j,k_1}}{\sqrt{\bar\theta}}-\frac{\theta_{j,k_2}}{\sqrt{\bar\theta}}}^2\vee \brac{\frac{1-\theta_{j,k_1}}{\sqrt{1-\bar\theta}}-\frac{1-\theta_{j,k_2}}{\sqrt{1-\bar\theta}}}^2}
\end{align*}
We have $\bar\delta=o(1)$ by assumption, and without loss of generality assume $\frac{1}{2}>\theta_{j,k_1}>\theta_{j,k_2}$ and denote $\theta:=\theta_2$, $\delta:=\theta_1-\theta_2=o(1)$. Then we have
\begin{align*}
    \frac{\brac{\theta_{j,k_1}-\theta_{j,k_2}}^2}{8\bar\sigma^2}=\frac{\delta^2}{2}\brac{1+o(1)}.
\end{align*}
On the other hand, we have
\begin{align*}
    -\log\brac{\sqrt{\theta_{j,k_1}\theta_{j,k_2}}+\sqrt{\brac{1-\theta_{j,k_1}}\brac{1-\theta_{j,k_2}}}}&=\frac{\brac{1+\bar C_{\theta+\delta,\theta}}}{2}\brac{\sqrt{\theta+\delta}-\sqrt{\theta}}^2\brac{1+o(1)}\\
    &=\delta^2\frac{\brac{1+\bar C_{\theta+\delta,\theta}}}{8\theta}\brac{1+o(1)}\\
    &=\frac{\delta^2}{8\theta\brac{1-\theta}}\brac{1+o(1)},
\end{align*}
where we've used the fact that
% \begin{align*}
    $\bar C_{\theta+\delta,\theta}=\frac{\theta}{1-\theta}\brac{1+o(1)}.$
% \end{align*}
This implies that
\begin{align*}
   \frac{-\log\brac{\sqrt{\theta_{j,k_1}\theta_{j,k_2}}+\sqrt{\brac{1-\theta_{j,k_1}}\brac{1-\theta_{j,k_2}}}}}{\brac{\theta_{j,k_1}-\theta_{j,k_2}}^2/\brac{8\bar\sigma^2}}=\frac{1}{4\theta\brac{1-\theta}}(1+o(1))>1
\end{align*}
where the last inequality holds due to $\theta<1/2$.

\section{Proofs of Lemmas}

\subsection{Proof of Lemma \ref{lem:init}}\label{sec:proof-lem-init}
To keep notation simple in the analysis, for any $\s\in[K]^N$  we define  
    \begin{align*}
         \theta_{j,k}\brac{\s}:=\frac{\sum_{i\in[N]} R_{i,j}\II\brac{s_i=k}}{\sum_{i\in[N]} \II\brac{s_i=k}}, \quad j\in[J],k\in[K].
    \end{align*}
The following result is needed, whose proof is deferred to \Cref{sec:proof-theta-err}.
\begin{lemma}\label{lem:theta-err}
     Suppose Assumption \ref{assump:dense}-\ref{assump:balanced} hold. For any $\gamma_0>0$,  there exists some absolute constants $C_1,C_2>0$ depending only on $\gamma_0$ such that with probability at least $1-2JKN^{-\brac{2\gamma_0+1}}$,
\begin{align}\label{eq:theta-err-1}
    \max_{s:\ell_0\brac{\s,\s^\star}\le {\alpha }/\brac{2K}}\max_{j\in[J],k\in[K]} \frac{\ab{\theta_{j,k}\brac{\s}-\theta_{j,k}\brac{\s^\star}}}{\ell_0\brac{\s,\s^\star}}\le  \frac{C_1 K}{\alpha}\brac{1+\bar\sigma\sqrt{\frac{K\log N}{\alpha N}}},
\end{align}
and
\begin{align}\label{eq:theta-err-2}
   \max_{j\in[J],k\in[K]} \ab{\theta_{j,k}\brac{\s^\star}-\theta_{j,k}}>C_2\bar\sigma\sqrt{\frac{K \log N}{\alpha N}}.
\end{align}
where $\ell_0\brac{\s,\s^\prime}:=\frac{1}{N}\sum_{i\in[N]}\II\brac{s_i\ne s_i^\prime}$. As a result, for any (possibly random) $\s\in[K]^N$ such that $\ell_0\brac{\s,\s^\star}\le {\alpha }/\brac{2K}$, we have 
\begin{align*}
   \max_{j\in[J],k\in[K]} \ab{\theta_{j,k}\brac{\s}-\theta_{j,k}}\le\ell_0\brac{\s,\s^\star}\cdot {\frac{ C_1K}{\alpha}}\brac{1+\bar\sigma\sqrt{\frac{K\log N}{\alpha N}}}+C_2\bar\sigma\sqrt{\frac{K \log N}{\alpha N}},
\end{align*}
with probability at least $1-2JKN^{-\brac{2\gamma_0+1}}$.
\end{lemma}
By \Cref{prop:spec-clustering} and Markov's inequality, we obtain that if 
\begin{align*}
   \min\ebrac{\frac{\Delta/\bar\sigma}{K\brac{1+\sqrt{\frac{J}{N}}}},  \frac{\lambda_K/\bar\sigma}{\brac{\sqrt{N}+\sqrt{J}}}}\rightarrow\infty,
\end{align*}
then for $m\in[2]$, with probability exceeding at least $1-\exp\brac{-\brac{1-c_0}\Delta^2/\brac{8\bar\sigma^2}}$ we have 
\begin{align*}
    \ell\brac{\wt \s^{(m)},\s^{\star}_{\calS_m}}\le \exp\brac{-\brac{1-c_0}\frac{\Delta^2}{8\bar\sigma^2}}.
\end{align*}
\sloppy where $\s^{\star}_{\calS_m}$ is sub-vector of $\s^\star$ with the index set $\calS_m$, and we've used that  $\exp\brac{-N/2}<\exp\brac{-\brac{1-c_0}\Delta^2/\brac{8\bar\sigma^2}}$. Without loss of generality, we assume $ \ell\brac{\wt \s^{(m)},\s^{\star}_{\calS_m}}= \ell_0\brac{\wt \s^{(m)},\s^{\star}_{\calS_m}}$. Combined with Lemma \ref{lem:theta-err}, we then have
\begin{align*}
   \max_{j\in[J],k\in[K]} \ab{\hat\theta^{(m)}_{j,k}-\theta_{j,k}}\le C_1\exp\brac{-\brac{1-c_0}\frac{\Delta^2}{8\bar\sigma^2}}+C_2\bar\sigma\sqrt{\frac{K \log N}{ N}}.
\end{align*}
with probability at least $1-\exp\brac{-\brac{1-c_0}\Delta^2/\brac{8\bar\sigma^2}}-\exp\brac{-C\sqrt{N/K}}-N^{-\brac{1+\gamma_0}}$, provided that there exists some universal constant $\gamma_0>0$ such that 
\begin{align*}
     JK \lesssim N^{\gamma_0},\quad N\gtrsim K\log N.
\end{align*}
The proof is completed by noticing that $I^\star\asymp \Delta^2/\bar\sigma^2$, $JK\le  \exp\brac{\brac{1-c_1}\Delta^2/\brac{8\bar\sigma^2}}$ for some $c_1\in (c_0,1)$, $\bar\sigma\asymp 1$ and
\begin{align*}
  \frac{\Delta}{{J^{1/2}}K^{3/4}\brac{\log N}^{1/4}/{N^{1/4}}}\rightarrow\infty.
\end{align*}

\subsection{Proof of Lemma \ref{lem:theta-err}}\label{sec:proof-theta-err}
We introduce the following notations:
\begin{align*}
    \bar\theta_{j,k}\brac{\s}:= \frac{\sum_{i\in[N]} \theta_{i,s_i^\star}\II\brac{s_i=k}}{\sum_{i\in[N]} \II\brac{s_i=k}},\quad \frakE_{j,k}(s):=\frac{\sum_{i\in[N]} E_{i,j}\II\brac{s_i=k}}{\sum_{i\in[N]} \II\brac{s_i=k}}.
\end{align*}
By definition we have $\bar\theta_{j,k}\brac{\s^\star}=\theta_{j,k}$ and $\frakE_{j,k}(s)=\theta_{j,k}\brac{\s}-\bar\theta_{j,k}\brac{\s}$, and hence
\begin{align}\label{eq:theta-err-decomp}
   & \ab{\theta_{j,k}\brac{\s}-\theta_{j,k}}\notag\\
    &= \ab{\theta_{j,k}\brac{\s}-\bar\theta_{j,k}\brac{\s}+\bar\theta_{j,k}\brac{\s}-\bar\theta_{j,k}\brac{\s^\star}+\bar\theta_{j,k}\brac{\s^\star}-\theta_{j,k}\brac{\s^\star}+\theta_{j,k}\brac{\s^\star}-\theta_{j,k}}\notag\\
    &\le \ab{\frakE_{j,k}(s)-\frakE_{j,k}\brac{\s^\star}}+\ab{\bar\theta_{j,k}\brac{\s}-\bar\theta_{j,k}\brac{\s^\star}}+\ab{\theta_{j,k}\brac{\s^\star}-\theta_{j,k}}.
\end{align}
We first bound the term $ \ab{\frakE_{j,k}(s)-\frakE_{j,k}\brac{\s^\star}}$. Notice that
\begin{align}\label{eq:frakE-decomp}
    &\ab{\frakE_{j,k}(s)-\frakE_{j,k}\brac{\s^\star}}\notag\\
    &\le \ab{\frac{\sum_{i\in[N]} E_{i,j}\II\brac{s_i=k}}{\sum_{i\in[N]} \II\brac{s_i=k}}-\frac{\sum_{i\in[N]} E_{i,j}\II\brac{s^\star_i=k}}{\sum_{i\in[N]} \II\brac{s_i=k}}}+\ab{\frac{\sum_{i\in[N]} E_{i,j}\II\brac{s^\star_i=k}}{\sum_{i\in[N]} \II\brac{s_i=k}}-\frac{\sum_{i\in[N]} E_{i,j}\II\brac{s^\star_i=k}}{\sum_{i\in[N]} \II\brac{s^\star_i=k}}}\notag\\
    &\le \frac{1}{\sum_{i\in[N]} \II\brac{s_i=k}}\brac{\ab{\sum_{i\in[N]}E_{i,j}\II\brac{s_i=k, s^\star_i\ne k}}+\ab{\sum_{i\in[N]}E_{i,j}\II\brac{s_i\ne k, s^\star_i= k}}}\notag\\
    &+\ab{\frac{\sum_{i\in[N]} \sqbrac{\II\brac{s^\star_i=k}-\II\brac{s_i=k}}}{\sum_{i\in[N]} \II\brac{s_i=k}}}\ab{\frac{\sum_{i\in[N]} E_{i,j}\II\brac{s^\star_i=k}}{\sum_{i\in[N]} \II\brac{s^\star_i=k}}}.
\end{align}
Note that the following fact holds for any $\s\in[K]^N$ such that $\ell\brac{\s,\s^\star}\le \alpha /\brac{2K}$:
\begin{align*}
   \sum_{i\in[N]} \II\brac{s_i=k}\ge \sum_{i\in[N]} \II\brac{s^\star_i=k}-\sum_{i\in[N]} \II\brac{s^\star_i\ne s_i}\ge \frac{\alpha N}{K}-\frac{\alpha N}{2K}\ge \frac{\alpha N}{2K}.
\end{align*}
% Combined with Lemma \ref{lem:sub-subset}, we obtain that
% \begin{align*}
%     &\frac{1}{\sum_{i\in[N]} \II\brac{s_i=k}}\brac{\ab{\sum_{i\in[N]}E_{i,j}\II\brac{s_i=k, s^\star_i\ne k}}+\ab{\sum_{i\in[N]}E_{i,j}\II\brac{s_i\ne k, s^\star_i= k}}}\\
%     &\le \frac{4K}{\alpha N}\brac{\sqrt{N\ell\brac{\s,\s^\star}}\max_{S\subset [N],\ab{S}\le \alpha N/(2K)}\ab{\frac{1}{\sqrt{\ab{S}}}\sum_{i\in S }E_{i,j}}}\\
%     &\lesssim\bar\sigma\sqrt{\alpha^{-1}K\ell\brac{\s,\s^\star}},
% \end{align*}
% where the last inequality holds with probability exceeding $1-\exp\brac{-C\sqrt{\alpha N/K}}$. 
Combined with the fact that $\ab{E_{i,j}}\le 1$, we thus obtain the following bound for the first term in \eqref{eq:frakE-decomp}:
\begin{align*}
    &\frac{1}{\sum_{i\in[N]} \II\brac{s_i=k}}\brac{\ab{\sum_{i\in[N]}E_{i,j}\II\brac{s_i=k, s^\star_i\ne k}}+\ab{\sum_{i\in[N]}E_{i,j}\II\brac{s_i\ne k, s^\star_i= k}}}\le \frac{4K}{\alpha }\ell\brac{\s,\s^\star}
\end{align*}
Next we bound the  second term in \eqref{eq:frakE-decomp}. By General Hoeffding's inequality, we obtain that 
\begin{align*}
    \PP\brac{\ab{\frac{\sum_{i\in[N]} E_{i,j}\II\brac{s^\star_i=k}}{\sum_{i\in[N]} \II\brac{s^\star_i=k}}}\ge t }\le 2\exp\brac{-\frac{ct^2}{\bar\sigma^2N_k}}.
\end{align*}
By choosing $t=\bar\sigma\sqrt{CN_k \log N}$ for some sufficiently large $C>0$, we have that 
\begin{align*}
    \ab{\frac{\sum_{i\in[N]} E_{i,j}\II\brac{s^\star_i=k}}{\sum_{i\in[N]} \II\brac{s^\star_i=k}}}\le C\bar\sigma\sqrt{\frac{K \log N}{\alpha N}},
\end{align*}
with probability exceeding $1-N^{-2\gamma_0-1}$ for some large constant $\gamma_0>0$. In addition, we have
\begin{align*}
    \ab{\frac{\sum_{i\in[N]} \sqbrac{\II\brac{s^\star_i=k}-\II\brac{s_i=k}}}{\sum_{i\in[N]} \II\brac{s_i=k}}}&\le \frac{2K}{\alpha N}\brac{\ab{\sum_{i\in[N]} \II\brac{s^\star_i=k,s_i\ne k}}+\ab{\sum_{i\in[N]} \II\brac{s^\star_i\ne k,s_i= k}}}\\
    &\lesssim \frac{K}{\alpha }\ell\brac{\s,\s^\star}.
\end{align*}
From \eqref{eq:frakE-decomp} and a union bound over $j\in[J]$ and $k\in[K]$, we  arrive at
\begin{align*}
    \max_{j\in[J],k\in[K]}\ab{\frakE_{j,k}(s)-\frakE_{j,k}\brac{\s^\star}}\lesssim {\alpha^{-1}K\ell\brac{\s,\s^\star}}\brac{1+\bar\sigma\sqrt{\frac{K\log N}{\alpha N}}}.
\end{align*}
with probability at least $1-JKN^{-2\gamma_0-1}$.

Next we bound the term $\ab{\bar\theta_{j,k}\brac{\s}-\bar\theta_{j,k}\brac{\s^\star}}$. Notice that by definition $\bar\theta_{j,k}\brac{\s^\star}=\theta_{j,k}$, hence
\begin{align*}
  \ab{\bar\theta_{j,k}\brac{\s}-\bar\theta_{j,k}\brac{\s^\star}} & =\ab{\frac{\sum_{i\in[N]}\sum_{k^\prime\in[K]} \brac{\theta_{i,k}-\theta_{i,k^\prime}}\II\brac{s_i=k,s_i^\star=k^\prime}}{\sum_{i\in[N]}  \II\brac{s_i=k}}}\le \frac{2K}{\alpha }\ell\brac{\s,\s^\star}.
\end{align*}
It suffices to bound $\ab{\theta_{j,k}\brac{\s^\star}-\theta_{j,k}}$ by 
\begin{align*}
   \max_{j\in[J],k\in[K]} \ab{\theta_{j,k}\brac{\s^\star}-\theta_{j,k}}= \max_{j\in[J],k\in[K]}\ab{\frac{\sum_{i\in[N]} E_{i,j}\II\brac{s^\star_i=k}}{\sum_{i\in[N]} \II\brac{s^\star_i=k}}} \le C\bar\sigma\sqrt{\frac{K \log N}{\alpha N}},
\end{align*}
with probability exceeding $1-JKN^{-2\gamma_0-1}$. 

Collecting \eqref{eq:theta-err-decomp} and the bounds for $ \ab{\frakE_{j,k}(s)-\frakE_{j,k}\brac{\s^\star}}$, $\ab{\bar\theta_{j,k}\brac{\s}-\bar\theta_{j,k}\brac{\s^\star}}$ and  $\ab{\theta_{j,k}\brac{\s^\star}-\theta_{j,k}}$, we can complete the proof.

\spacingset{1}
\bibliographystyle{apalike}
\bibliography{references}

\end{document}